\documentclass[]{aa}
\usepackage{graphicx}
\usepackage{txfonts}                                       
\usepackage{natbib}

\begin{document}
\title{Bar instability in cosmological halos}

\author{Anna Curir\inst{1}
 \and Paola Mazzei\inst{2}
  \and Giuseppe Murante\inst{1}}
\offprints{ A. Curir}
\institute{INAF-Osservatorio Astronomico di Torino. Strada Osservatorio 20
 -10025 Pino Torinese (Torino). Italy.   e-mail:curir@to.astro.it
\and 
     INAF-Osservatorio Astronomico di Padova. Vicolo Osservatorio 5 - 35122
 Padova. Italy
}
 

\date{Received/Accepted}
\abstract
{}
{We want to investigate the growth of bar instability in stellar disks embedded 
in a suitable dark matter halo evolving in a fully consistent cosmological
framework.} 
{We perform seven cosmological simulations to emphasise 
the role of both the disk-to-halo mass ratio
and of  the Toomre parameter, Q, on the evolution of the disk.We also 
compare our fully cosmological cases with corresponding isolated
simulations where the same  halo, is extracted from the cosmological 
scenario and  evolved in physical coordinates.}
{A long living bar, lasting about 10 Gyr, appears in all our 
simulations. In particular, disks expected to be stable according to 
classical criteria,  form indeed weak bars.
We argue that such a result is due to the dynamical properties of our 
cosmological halo which is far from
stability and isotropy, typical of the classical halos used in
literature; it is  dynamically active, endowed of substructures
and infall.}
{At least  for mild self-gravitating disks, the study of the
bar instability using isolated isotropic halos, in gravitational equilibrium, 
 can lead to misleading results.
Furthermore, the  cosmological framework is needed for 
quantitatively investigating such an instability.}
\titlerunning{Bars in cosmology}
\authorrunning{Curir  Mazzei \& Murante}
\maketitle

\section{Introduction}
 Several work on bar instability in stellar disks   has been done using  N-body spherical halos (\citet{Sel81}; \citet{atha87}; \citet{Deb00}; \citet{Pat00}; \citet{May04}).\\ 
\citet{Cu99} were the first to emphasise
the  role of  both the geometry and of the dynamical state of a 
live dark matter (DM) halo in 
enhancing the bar  formation.  Progressive efforts
for improving  models of the halo  have been made in
the last years (\citet{Ma01}; \citet{athami02}), 
taking into account also the informations coming from the cosmological hierarchical
clustering scenario of structures formation about  density distribution and 
concentration of DM halos. 
In the meanwhile, the ever-growing computing power available to the community
has made possible to start simulations of formation and evolution of 
galaxies in a fully cosmological context. 
First  works devoted to deep our understanding of disk galaxies in such a
scenario (\citet{Gov02}; \citet{Aba03}) have shown that  is very difficult to obtain pure 
disk galaxies mainly because of the
high angular momentum loss of the gaseous component. Even with 
a careful choice of the hosting DM halo, the simulated
galaxies appear to have over--massive bulges compared to their disks.
In a  recent paper, \citet{Spri04} claim to have overcome
most problems, however the bar instability has not yet  been 
analysed in a cosmological framework.
Furthermore, the high CPU cost of such simulations does not yet
allow to explore the role of  several  
parameters, most related to the phenomenological treatment of the 
star formation rate and
feedbacks,  on the morphologies of the generated
 galaxies (\citet{MaCu03}; \citet{Ma03}).\\
In this work we  present the first attempt to analyse the growth of bar
instability in a fully consistent cosmological scenario.
We embed a pure stellar disk inside a  cosmological halo selected in a
suitable slice of Universe and follow its  evolution inside a cosmological
framework.
We want to explore how the bar instability  behaves 
and what is the role of such a scenario. In particular we want to address,
 besides the role played by the disk-to-halo mass ratio, that of
the dynamical state of  such an halo as given by 
its substructure and  infall, or more generally by its evolution. 
Our model cannot be
viewed as a general, {}``all-purpose{}'' galaxy evolution model, since the
\emph{gradual} formation and growth of the stellar disk is  a fundamental
component of the galaxy evolution itself. However
our approach allows  to vary parameters like the disk-to-halo mass ratio 
and the disk temperature, as given by {\it Q} parameter, 
to analyse the growth of the bar instability and its dependence on such
parameters for the first time in a self-consistent cosmological framework.
We   analyse further  the
influence of the cosmological environment by comparing these results 
with those in an isolated scenario with the same halo.\\
The plan of the paper is the following:  Section 2 and 3 describe technical
details, in particular the  recipe for the initial $disk+halo$ system and our  framework,
focusing on the cosmological evolution  and on the properties of the halo.
In Section 4 we present the whole set of our disk+halo
simulations; in Section 5 we point
out our  results in the cosmological context and the comparison with isolate runs.  In Section 6 is our discussion 
and in Section 7 our conclusions. 
In the Appendix we analyse
the robustness of our results, checking for particle 
resolution and softening length effects. 
 \section{Numerical method}
Our galaxy model consists of a truncated exponential disk \citep{Cu99}, self-consistently
embedded in a suitable DM halo extracted from a cosmological simulation.
To select the DM halo, we perform a low-resolution (128\( ^{3} \) particles)
simulation of a {}``concordance{}'' \( \Lambda  \)CDM cosmological model:
\( \Omega _{m} \) =0.3, \( \Omega _{\Lambda } \)=0.7, $\sigma_8 = 0.9$, 
\( h \)=0.7, where \( \Omega _{m} \) is
the total matter of the Universe, \( \Omega _{\Lambda } \) the
cosmological constant, $\sigma_8$  the normalisation of the power spectrum,  
and $ h $ the value of the Hubble constant in units of 
$100 \, h^{-1}\,$ km\, s$^{-1}$\, Mpc$^{-1}$. 
The
box size of our simulation is $ 25 h^{-1}$ Mpc, which  allows us
an adequate cosmological tidal field and no boundary effects
on our disk. The initial redshift is 20.
We employ the public parallel N-body treecode GADGET \citep{Spri01}.
Our initial condition code has been adapted from the setup code of ART
(\citet{Kra97}; \citet{Kra99}; courtesy of A. Klypin).\\
From this simulation
we identify  the DM halos  at {\it z}=0  in the mass 
 \footnote{In the following, we will  refer to the mass as
the virial mass i.e. that enclosed in a sphere with overdensity
$ \delta = \rho /\rho _{crit}=178\cdot \Omega _{m}^{0.44}$ 
\citep{Nav00}.} range
0.5- 5\( \cdot  \)10\( ^{11} h ^{-1} \) M\( _{\odot },\) with a standard  
friends-of-friends algorithm.
We discard the halos belonging or near to
overdense regions (see Sect. 3). Then we follow back 
the simulation and discard  those which suffer significant
mergers  after a redshift of \( \sim  \)5. 
So we select one suitable DM halo with a mass
 M\( \sim  \)10\( ^{11} \)\( h^{-1} \) M\( _{\odot } \) (at {\it z}=0). We 
resample it with the multi-mass technique 
described in \citet{Kly01}. The particles of the DM halo, and those
belonging to a sphere with a radius $4 h^{-1}$ Mpc,  are followed
to their Lagrangian position and resampled to an equivalent resolution of 
1024 \(^{3} \) particles. 
The total number of DM particles in the high resolution region
is $1216512$ which corresponds to a DM  mass resolution of
\( 1.21\cdot 10^{6} h^{-1}\)M\( _{\odot } \). 
The needed high frequency power is added without modifying
the low-frequency Fourier phases of the CDM power spectrum
in our low resolution run. The high resolution zone is surrounded by three
shells with lower and lower resolution, the lowest one including all the 
remaining (not resampled) particles among the initial 128\( ^{3} \) set.\\ 
The size of the initial Lagrangian region is large enough to resolve with high
resolution  not only the DM halo, but also its accreting
sub--halos.
The high-resolution DM halo is followed 
to the redshift {\it z}=0. We checked that \emph{no}  lower 
resolution particles (intruders) are ever present at a radius lower than 
 \( \sim  \) 2 \( h^{-1} \)Mpc from its centre.  This corresponds 
to the particle with the minimum gravitational energy.\\
Our approach  allows
us to account for the cosmological tidal field acting on the DM halo
and to accurately follow the evolution of the selected halo in a self-consistent 
way.\\
We carried out two sets of simulations embedding the 
galactic disk in the halo at the redshifts $z=2$ and $z=1$ respectively. 
The first choice corresponds to  10.24 Gyr down to $z=0$ in our chosen cosmology, the second one to 
7.71 Gyr.\\
Details of our model disk are presented elsewhere \citep[e.g.][]{Cu99}. Here we
summarise the main features of the disk.
The spatial distribution of the star particles  follows the
exponential surface density law: \( \rho _{stars}=\rho _{0}\exp -(r/r_{0}) \)
where \( r_{0} \) is the disk scale length, \( r_{0}=4h^{-1} \)kpc, and \( \rho _{0} \)
is the surface central density. The disk is truncated at five scale lengths with
a radius:
\( R_{disk}=20h^{-1} \)kpc. To obtain each disk particle's position according
to the assumed density distribution, we used the rejection method \citep{Pre86}.

The vertical coordinate is extracted from a Gaussian distribution with a standard
deviation equal to 1\% of the disk radius. Circular velocities are assigned
analytically to disk particles accounting for the global (disk+cosmological halo) potential,
$\Phi$.
The radial velocity dispersion ${\sigma}_R$  is assigned through a Toomre
parameter {\it Q}. {\it Q}  is
initially constant at all disk radii and is it defined as 
${Q}= {{ {\sigma}_R \,\kappa } \over{ 3.36 \,G \,\Sigma }}$, where 
 $\kappa$ is the epicyclic
frequency, and ${\Sigma}$  the surface density of the disk. 
According to the isothermal sheet approximation, the ratio of radial to
vertical dispersion is fixed and constant through the disk, moreover the azimuthal dispersion is linked to the radial dispersion via the
epicyclic approximation \citep{hern93}.
The final velocity distributions are Gaussian,
with the dispersions given above.\\ 
 Assigning  a constant
initial {\it Q}, we can  classify easily our disks on the basis of
the initial
temperature.
We explore  two values of {\it Q}: 1.5, which corresponds to a {\it  warm}
disk, and 0.5, to a {\it cold} disk.
The  average {\it Q} value of stars in the Milky Way is estimated 
between 1 and 3 \citet{bintre87}, however the evolution of such 
a parameter starting from high {\it z} is not known. \\
Our model of galaxy is very simplified. Neither gas nor star 
formation are introduced since we aim  to focus on the 
\emph{gravitational} effect of the halo on the disk and to have hints on the
\emph{gravitational}  feedback of the disk itself on the halo.
Moreover,
our technique is such that the CPU cost of one simulation, while large, is still 
much lower than the cost of galaxy formation simulation like 
that by \citet{Aba03},
even if our force and mass resolution are comparable. Thus our work 
could give insights  into   self--consistent galaxy formation scenario.\\
In the following we summarise the main steps of our approach:\\
i) the halo is identified at redshift $z=0$ ;\\
ii) its particles are tracked back to the selected redshift 
(i.e. $z=1$ and $z=2$), and the minimum of their potential well is calculated;\\
iii) a \emph{sphere} of radius R\( _{sphere} \)=3R\( _{disk} \) is extracted from
the high resolution simulation; its bulk velocity and the position of
its centre are recorded. R\( _{sphere} \) is chosen to ease the comparison of our results with previous numerical work on disk stability, e.g. \citet{Cu99} 
and \citet{Ma01}.
Note that $ R _{sphere}  \geq  R _{vir} $ and
$ M_{halo}^{vir} \)\( \geq M_{halo}^{sphere}$;\\ 
iv) the vector angular momentum \( \overrightarrow{J} \) and the gravitational
potential \( \Phi  \) are calculated, inside $R_{sphere}$;\\
v) the disk, in gravitational equilibrium with the potential \( \Phi  \)
and rotating in a plane perpendicular to \( \overrightarrow{J} \), is generated;\\
vi) we embed the disk in the high resolution simulation, at the chosen redshift,
with its centre of mass in the minimum potential well of the DM halo;\\
vii) the bulk velocity of the halo is added to the star particles.\\
The cosmological simulation is evolved then,
in comoving coordinates, to the final redshift, $z=0$.

\section{The DM halo}
After selecting the halo and resampling at the higher resolution
the corresponding Lagrangian region, we run  the DM simulation, to extract 
the halo properties in absence of any embedded stellar disk.
The mass of our  halo at $z=0$,
\(1.03\cdot 10^{11}h^{-1} \) M\( _{\odot } \), corresponds to a
 radius,  \( R_{vir} = 94.7h^{-1}\)kpc, which entails
84720 halo particles.
The nearest DM halo
\footnote{ Halos have been identified
using the friends of friends algorithm with a linking length $l = 0.15$, mean
interparticle distances, with more than 8 particles.}
 more massive than $10^{10} h^{-1} M_\odot$ is 
$\sim 1900 h^{-1}$ kpc away from the centre of our halo;
 the less massive one,  having mass of $4.6\cdot 10^7 h^{-1} M_{\odot}$,
is $\sim 215 h^{-1}$ kpc away. Moreover, the behaviour of
the density contrast, $\delta$, is
monotonically decreasing with the radius, and  $\delta$ falls below the
unity value at 
$\sim ~550h^{-1}, ~450h^{-1}, ~350 h^{-1}$ physical kpc away from the centre of our
halo at $z=0, z=1$, and $z=2$ respectively.
Therefore, we conclude that the selected halo is  living in an 
under-dense environment.
\begin{figure*}
\centering
\includegraphics[angle=-90,width=10cm]{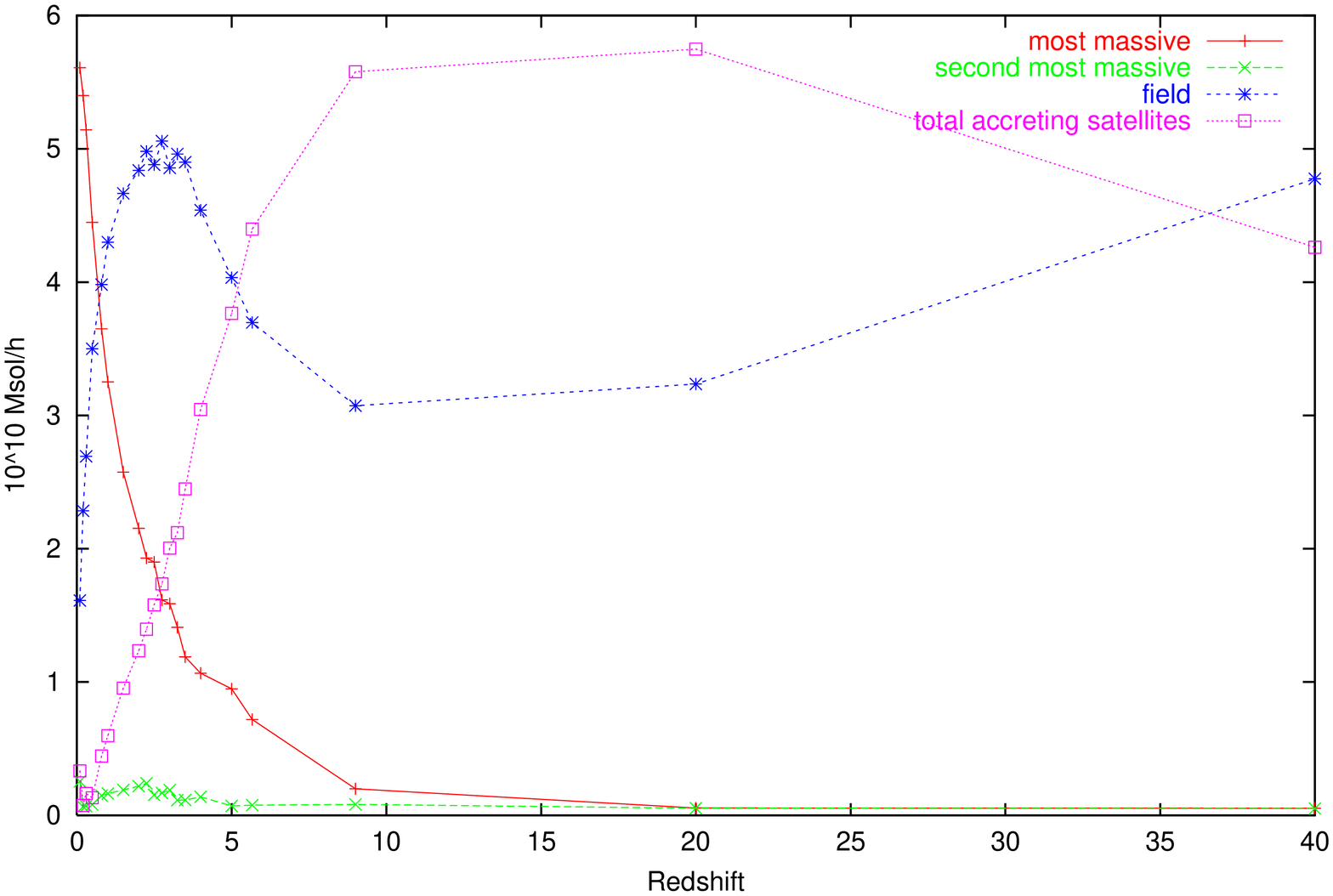}
\includegraphics[angle=-90,width=10cm] {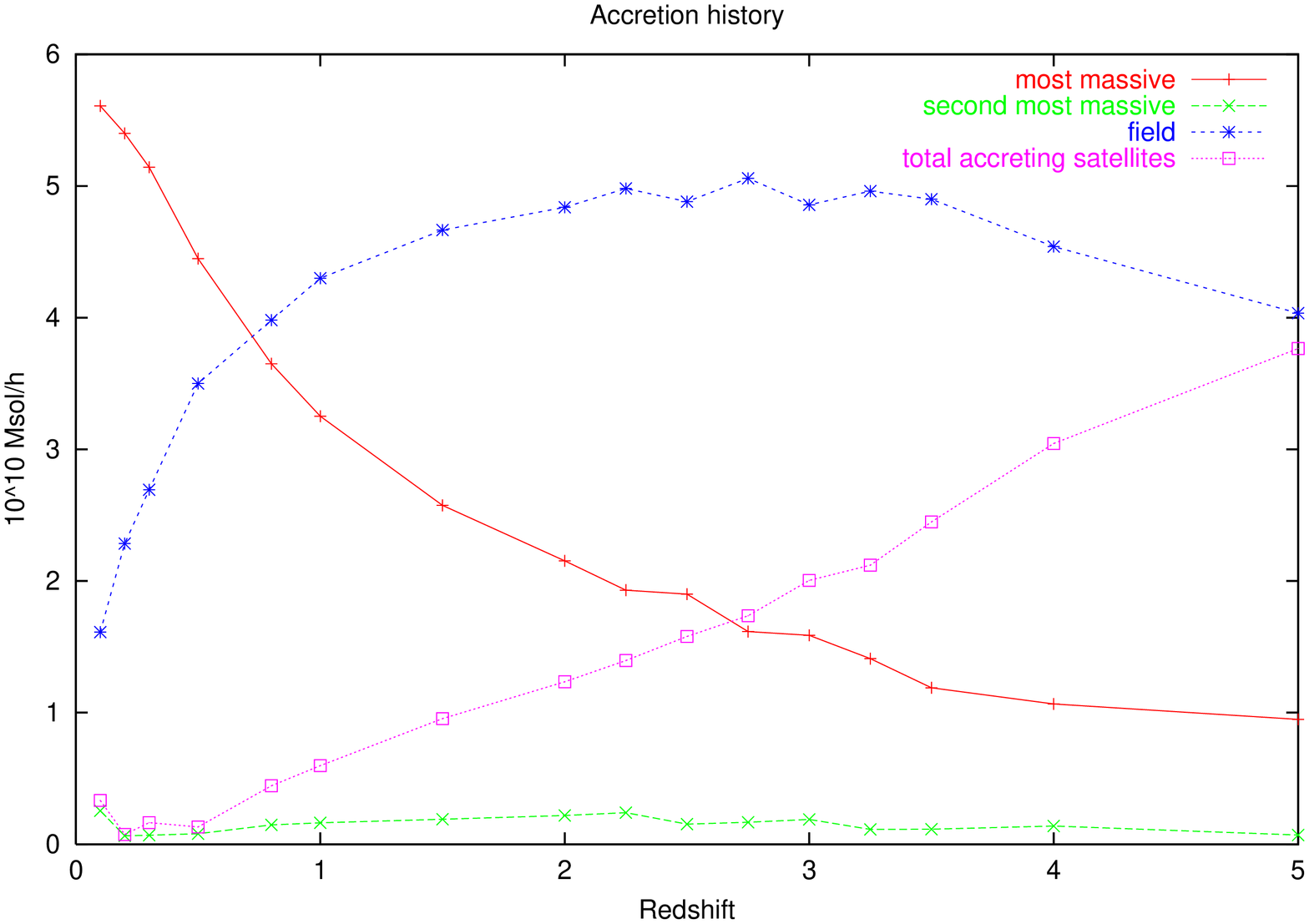}
\caption{Accretion history of our selected DM halo as a function of the redshift: top panel from redshift $z=40$ to $z=0$, bottom panel from $z=5$ 
to $z=0$;
solid (red) line shows the mass of the most massive progenitor of the halo,
long-dashed (green) line
that of the second most massive progenitor, dotted (magenta) line 
the total mass of progenitors (the most massive  excluded),
and short-dashed (blue) line the total mass of  {\it field } particles.}
\label{accrhist}
\end{figure*}
\begin{table*}
\caption{  DM halo properties}
\label{halotable}
\begin{tabular}{c c c c c c c c }
\hline\hline
 $z$ & $C_{vir}$ & $M_{vir}$ & $R_{vir}$ & $N_{part}$ & $\lambda$ & $\tau_{60}$ &  

 $\tau_{20}$\\  
\hline
 0 & 18.0 & 1.03 $\cdot 10^{11}$ & 94.7 & 84720 & 0.01 & 0.60 & 0.80 \\
 1 & 13.2 & 7.7  $\cdot 10^{10}$ & 51.6 & 63886 & 0.02 & 0.78 & 0.92 \\
 2 &  8.3 & 5.2  $\cdot 10^{10}$ & 30.9 & 42660 & 0.04 &  0.92 & 0.90 \\
\hline
\end{tabular}
\\
I   col: redshift\\
II  col: concentration parameter $C_{vir}$\\
III col: virial mass, in $h^{-1} M_{\odot} $\\
IV  col: virial radius, in $h^{-1}$ physical kpc \\
V  col: number of DM particles within the virial radius\\
VI   col: spin parameter\\
VII  col: triaxiality parameter within a sphere of radius $60 h^{-1}$physical kpc\\
VIII col: triaxiality parameter within a sphere of radius $20 h^{-1}$physical kpc \\
\end{table*}

The accretion history of our halo has been
calculated as follows. Starting from redshift $z_1=0$, we identified DM halos
using the public halo finder SKID 
\footnote{http://www-hpcc.astro.washington.edu/tools/skid.html}
\citep{Sta01} at the redshift $z_2=z_1+dz_1$, 
corresponding to our previous simulation output. 
We then define as {\it progenitor} of our  halo a SKID group at the redshift 
$z_2$, if at least a fraction $f=30$\%  of its particles  come  from the halo at the   redshift $z_1$. We also identify as {\it accreting field particles} 
all the DM 
particles not belonging to any SKID group but belonging to the halo. We then 
iterate the procedure, using the simulation output corresponding to the 
redshift $z_3=z_2+dz_2$; the progenitors are now all the groups which have at least a 
fraction $f$ of particles coming from progenitors at $z_1$ {\it or} from the accreting 
field particles, and so on for the earlier redshifts. We check that the 
qualitative behaviour of the accretion history is not dependent on the 
value of $f$ (we also tested $f$=20\% and $f$=50\%) and on the parameters used in 
SKID (we use a typical object size $\tau = 5 h^{-1}$kpc but also  the effect of
3$h^{-1}$\,kpc and $6h^{-1}$\,kpc have been explored).  
 From Fig.  \ref{accrhist}, we can note that the halo suffers its
last major merger (i.e. a merger between two progenitors whose masses have a
ratio which is not larger than 3)  at  $z=9$. 
After $z \sim 5$,
the most important contribution to its mass comes from accreting field 
particles. 
This contribution declines after $z \sim 2$ 
becoming less and less important. At $z \sim 0.9$, the total accreting mass
is smaller than the mass of the larger sub--halo. 
Thus we  conclude that or  halo has no significant merger during 
the time hosting our stellar disk, nor immediately before.\\
The properties of the selected halo at three relevant redshifts are listen in 
Table \ref{halotable}.
Its density profile  is well--fitted by a NFW form (\citet{Nav96}; \citet{Nav97})
at $z \leq 2$. The concentration, $C_{vir}$
\footnote{ We note however that $C_{NFW}$
is defined  against $R_{200}$, the radius enclosing a sphere with overdensity equal to
200 times the critical density of the Universe, and not against $R_{vir}$ as here;
therefore, in our cosmological model, it is always $C_{NFW}<C_{vir}$.
At $z=0$ our halo has $C_{NFW} \sim 14$.} 
here defined as 
$R_{vir}/R_s$, takes an high value, 18.1,  confirming that this halo does ``form'' 
at  quite high redshift  (see \citet[e.g.][]{Wec02} for a discussion about
the link between  concentration and  assembly history of the halo).
The dimensionless spin parameter of the halo is  defined as: 
 $\lambda = { J \over \sqrt{2} MVR }$
\citep{bul01}
where $J$ is the angular momentum inside a sphere of radius $R$ and $V$ is the halo 
circular velocity, $V^2=GM/ R$. Its values in Table \ref{halotable} 
are near to
the average ones for our cosmological model 
\citep[$\lambda = 0.035$;][]{Mal02} 
\section{Disk simulations}
We performed seven simulations of the disk+halo system 
as described here below (Sect. 4.1).
By comparing results of such a set of 7 simulations (Sect. 5) with the 
DM-only, we
disentangle the effect of the stellar disk on the halo evolution in the
cosmological framework.
Several simulations of
the disk+halo isolated system are also run, to disentangle the effect of the 
cosmological environment (Sect. 4.2 and 5.2).
We used 56000 star particles to describe our disk;
the (Plummer-equivalent) softening length, the same for DM and
star particles, is 0.5$\,h^{-1}\,$kpc in comoving
coordinates \footnote{Note that, since the disk is modelled in physical
  coordinates and embedded in the cosmological halo at redshifts  $z=2$ and
  $z=1$, its thickness is  larger than the value of the Plummer
  softening we use.}.
We used a time-step criterion based on the local dynamical time (criterion {}``3{}''
of the GADGET code), which provides  $2-6 \times 10^4$ time-steps from
$z=2$ to $z=0$ (except one case, simulation 5 of Table 2
which needs only $\sim 7000$ time steps). 
The most CPU--expensive of our simulations needed $\sim 5000$ CPU
hours to be completed on the SP4 computer (CINECA computing center).
\subsection{Cosmological cases}     
\begin{table*}
\caption{ Simulations: initial values.}
\label{cosmsimtable}
\begin{tabular}{c c c c c c c c c}
\hline\hline
 $N$ & $Q$ & $M_{disk}$ & $ {\it z} $ & $M_{DM}$ & $R_{DM}$ & $ \alpha \, r_m$ &
${v_m}\over{{(\alpha G M_{disk})}^{1/2}}$ &  halo \\  
\hline
 c1 & 0.5 & 1 & 2 & 0.64 & 0.64 & 1.9 & 0.67 & ${}$ \\
 c2 & 0.5 & 0.33 & 2 & 0.64 & 1.94 & 1. & 1.08 & ${}$ \\
 c3 & 0.5 & 0.1 &  2 & 0.64 & 6.4 & 0.9 & 1.68 & ${}$ \\
 c4 & 1.5 & 0.33& 2 & 0.64 & 1.94 & 1. & 1.08 & ${}$ \\
 c5 & 1.5 & 0.1 & 2 & 0.64 & 6.4  & 0.9 & 1.68 & ${}$ \\
 c6 & 0.5 & 0.33 & 1 & 0.67  & 2.0 & 1.05 & 1.05 & ${}$\\
 c7 & 0.5 & 0.1 & 1 & 0.67 & 6.7 &  1. & 1.6 & ${}$ \\ 
 i1 & 1.5 & 0.33 & ${}$ & 0.64 & 1.94 & 1 & 1.08 & cosm\\
 i2 & 1.5 & 0.1 &  ${}$ &  0.64 & 6.4 & 0.9 & 1.68 & cosm\\
 i3  & 1.5 & 0.33 & ${}$ & 0.64 & 1.94 & 1 & 1.08 & cosm/frozen disk\\
 i4 & 1.5 & 0.33 & $ {} $ & 0.95 & 2.87 & 0.85 & 1.5 & NFW \\
 i5 & 1.5 & 0.1  & ${} $ & 0.95 & 9.5&  1.27 & 1.25 & NFW \\
\hline 
\end{tabular}
\\
I   col: simulation number and simulation type (c: cosmological simulations,
i: isolated simulations) \\
II  col: {\it Q} initial value of the disk\\
III col: mass of the disk in $5.9\times 10^{10}\, M\odot$\\
IV  col: initial redshift  (for the cosmological cases)\\
V  col: initial DM mass inside the disk radius\\
VI   col: initial halo-to-disk mass ratio inside the disk radius\\
VII and VIII cols: \citet{Efs82} parameters, where: $\alpha={r_0}^{-1}$, $v_m$ is
the maximum rotational velocity, and $r_m$ the corresponding radius.\\
IX col: type of halo used (for the isolated cases) 
\end{table*}
\begin{table*}
\caption{ Simulations: final results}
\label{cosmsimtable_fin}

\begin{tabular}{c c c c c c c c }
\hline\hline
 $N$ & $M_{DM}$ & $R_{DM}$ &  $S_m$ & $Q_t$  & $a_{max}$ &  bulge & bars in bars\\  
\hline
 c1 & 0.79 & 0.8 & 0.42 & 0.38 & 7 & y & n\\
 c2 & 0.77 & 2.39 & 0.33 & 0.44 & 8 & y & n\\
 c3 & 0.73 & 7.41 & 0.8 & 0.07 & 3.8 & n & y\\
 c4 & 0.78 & 2.40 & 0.48 & 0.37 & 5 & weak & n\\
 c5 & 0.73 & 7.41 & 0.70 & 0.08 & 6.5 & n & y\\
 c6 & 0.79 & 2.43 & 0.35 & 0.42 & 6.8 & weak & n\\
 c7 & 0.77 & 7.73 & 0.58 & 0.16 & 5.0 & n & y\\
 i1 & 0.73 & 2.21 & 0.25 & 0.4  & 9.5 & y & n \\
 i2 & 0.51  & 5.1 & 0.68 & 0.1  & 8   & n & y \\
 i3 & 0.7  &  2.12 & 0.3 & 0.39 & 10  & y & n \\
 i4 & 1.   &  3.03 & 0.33 & 0.42 &  6 & y & n \\
 15 & 0.95 &  9.5  & 0    & 0   &  0 &  n & no bar \\
\hline
\end{tabular}
\\
I   col: simulation number and simulation type\\
II  col: DM mass inside the disk radius in $5.9\times 10^{10}\, M\odot$\\
III   col: halo-to disk mass ratio  inside the disk radius\\
IV col: maximum bar strength at $z=0$: strong
bar \citet{Ma01} require $S_m\le 0.6$\\
V col: bar strength evaluated according to  \citet{Comb81}, stronger bar
corresponds to higher values of $Q_t$\\
VI col: major axis (physical kpc) corresponding to the maximum bar strength\\
VII col: morphology of the inner region of the disk\\
VIII col: peculiar features inside the disk 
\end{table*}
The main parameters and the initial properties of this set of  simulations
 are listed 
in Table  \ref{cosmsimtable}.\\
A global stability criteria for bar instability in a disk galaxy
 is the one analysed in  \citet{Efs82}. In such paper the parameters  $\alpha
 \, r_m$ and  $v_m\over{{(\alpha M G)}^{1/2}} $ (where  $v_m$ is the maximum
 value of the disk rotational curve,  $r_m$ the corresponding radius,
 ${\alpha} = {{r_0} ^{-1}}$ and $M$ is the disk mass) have been defined.
\citet{Efs82} stated the criterion  ${v_m\over{{(\alpha M G)}^{1/2}}}  \geq
      {1.1} $ over the range  $0.1 \leq {\alpha \, r_m} \leq 1.3$ for a disk
      model being stable to bar formation. The values of these parameters
      are reported in Table  \ref{cosmsimtable}.\\
Simulations c1, c2 and c3 in Table \ref{cosmsimtable} refer to a {\it cold} disk 
($Q=0.5$). In simulation c1, at the final time (i.e. $z=0$) 
the baryon fraction inside $R_{vir}$, 
$f_{b}=M_{disk}/M_{disk+DM} \sim 0.34$,  is
44\% less than its initial value, 0.53. 
The final baryon fraction of simulation c2 is $\simeq 0.16$, compared
with its initial value, 0.28.
Simulation c3 provides $f_{b}\simeq 0.05$ at $z=0$, 50\% less than its initial value.
Simulations c4 and c5
provide the same final baryon's fractions as the corresponding simulations with
the lower  Toomre's parameter.\\
Simulation c6 and c7, which explore the  
role of the initial redshift on the bar instability, provide quite the same
final values of the baryon's fraction as 
the corresponding simulations c2 and c3.
Neither the Toomre  parameter  nor the initial redshift affect
the evolution of this ratio which is driven by the mass of the stellar disk.  
While the baryon fraction of simulation c1 
is  too high to be consistent with the 
cosmological value 0.166 \citep{Ett03}, all the other simulations give baryon
fractions in the allowed range. We however emphasise that the aim of the current work 
is not to build a realistic galaxy model, but to study the effect of different
halo-to-disk mass ratios on the onset of the bar instability. 
We verify that the inclusion of the disk
does not result in significant changes in the accretion history of the DM halo.\\
\subsection{Isolated cases}
We also performed several simulations of the isolated  disk+halo system using
the same halo  as extracted from our cosmological simulations at $z=2$ (Sect. 5.2
and Appendix).
By comparing results of this set of simulations with the previous ones
we aim to disentangle the effect  of the large scale cosmological structure
and of cosmological expansion on the system evolution. 
Moreover, such results  are directly comparable both with
our previous works (\citet{Cu99}; \citet{ Ma01}) 
and with those in literature
\citep{atha87}. 
The initial and final values for these simulations are listed in Tables
\ref{cosmsimtable} and \ref{cosmsimtable_fin}.  We stress that our isolated
halo is produced by a non dissipative collapse in a cosmological scenario.
As a consequence its mass distribution is not spherically symmetric. Moreover
such a halo is anisotropic and endowed with a spin parameter and substructure.
Therefore it is different from the standard isolated halos used in literature
to study bar instabilities, since it keeps a relic cosmological signature.
When the halo is extracted  from its cosmological environment, the large scale
structure, the continuing matter infall and the expansion of the Universe no 
longer influence its evolution. Such a halo {\it cannot}  be in gravitational
equilibrium, because its evolution has not yet completed neither at $z=2$ nor
at $z=1$, as shown in Fig.  \ref{accrhist}. For this reason, the results
presented below, concerning the behaviour of the disk embedded in such a
``isolated'' halo have to be compared with similar cases, in which 
non--equilibrium DM halos are used, as e. g. in  \citet{Cu99}; \citet{ Ma01}.
After subtracting the CM velocity and embedding the disk, as
described in Sect. 2  items i-v, we integrated the
system in {\it physical} coordinates (the effect of the cosmological expansion
  is therefore ruled out in these models) .  A further
difference is that the softening length is now in physical units. 
We have at least 10000 time steps
from the initial time to $t=10.24$ Gyr corresponding to redshift 0.\\
Finally, in order to disentangle the effect of the geometry and of the spin of an
isolated halo we also performed two simulations using a Navarro Frenk and
White (NFW) halo having the same virial radius and mass as our cosmological
one. The initial and final values of these two simulations are listed in the
two last lines of Tables \ref{cosmsimtable} and \ref{cosmsimtable_fin}. \\ 
\section{Results}
In this section we  present  the evolution of 
isodensity contours of the different cosmological simulations. From these
contours we evaluate the final bar strengths  which are reported in Table
\ref{cosmsimtable_fin}. Spatial resolution of the maps is always 
0.5$h^{-1}$ {\it physical \,} kpc and the
box size is 40 times the spatial resolution.
Contours are computed at 11 fixed levels ranging from  
$2\times 10^{-4}$  to $0.015$  in term of  fraction of stars/spatial resolution
within the total number--density of stars in the map. 
Following \citet{Cu99}, we define,  as a measure of the bar strength, the
maximum value of 
the  axial ratio, $S_m=b/a$ (Table \ref{cosmsimtable_fin}): 
a strong bar corresponds to  $S_m\leq0.6$ or to an
ellipticity,  $\epsilon=(1-b/a)$, larger than 0.4.
\subsection{Morphologies of the stellar disk in the cosmological framework}
\begin{figure*}
\centering
\includegraphics[width=10cm]{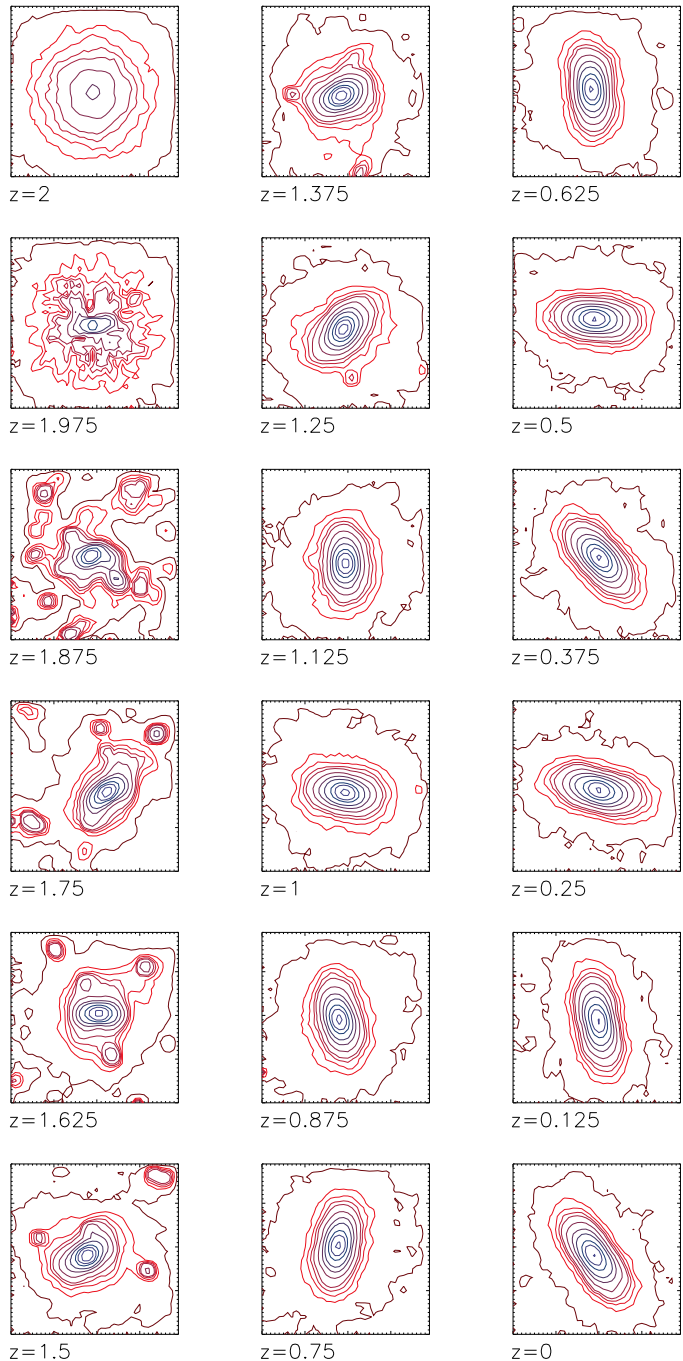}
\caption{ Evolution of  isodensity contours of simulation c1 
at 11 fixed levels (see Sect. 5).  The size of all the frames 
is $20 h^{-1}$  physical kpc, here and in all the following figures in which
isodensity contours are shown.}
\label{dens1}
\end{figure*}

\begin{figure*}
\centering
\includegraphics[width=10cm]{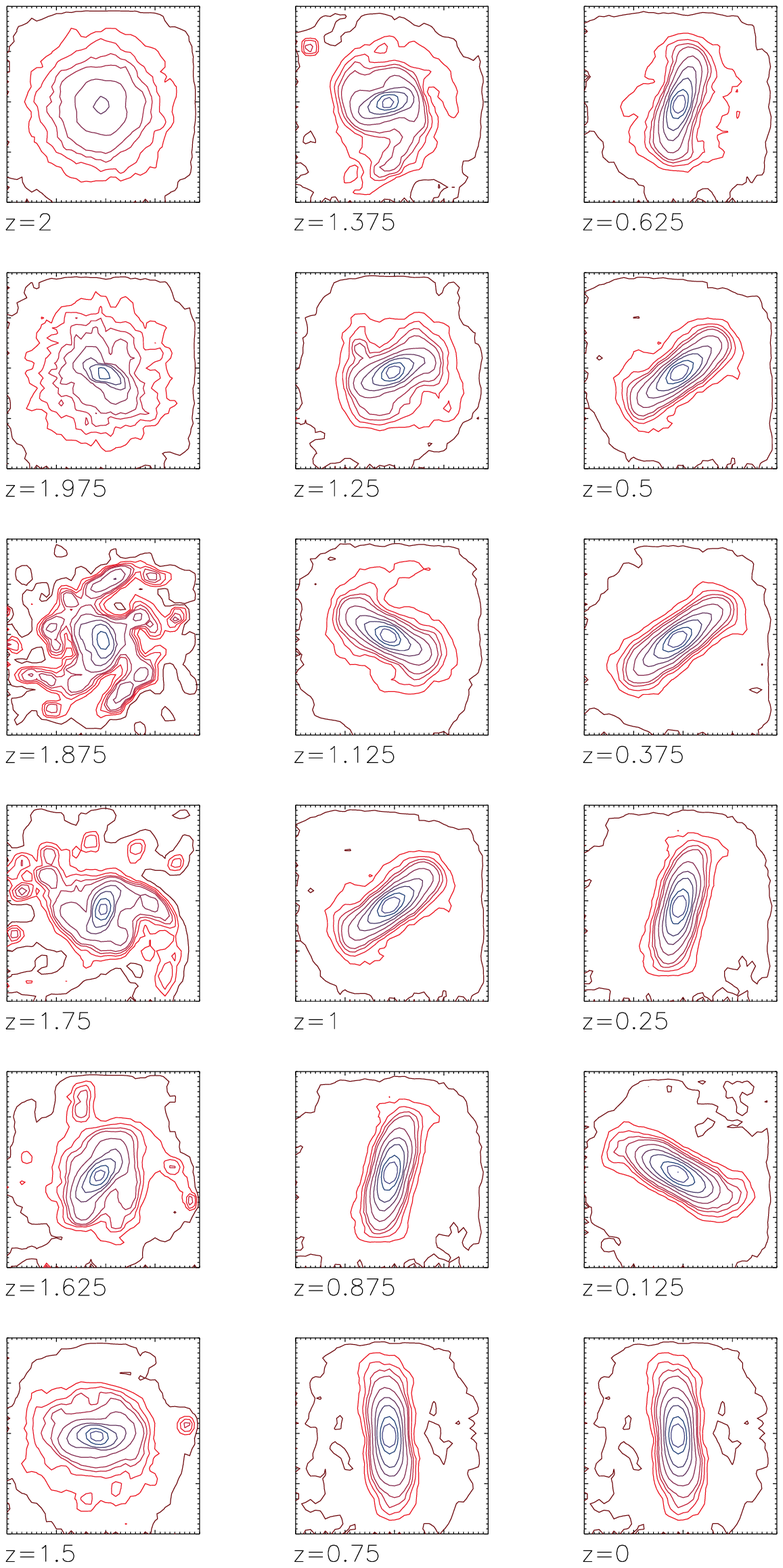}
\caption{Evolution of density contours of simulation c2 as described in Fig. \ref{dens1}.}
\label{dens2}
\end{figure*}

\begin{figure*}
\centering
\includegraphics[width=10cm]{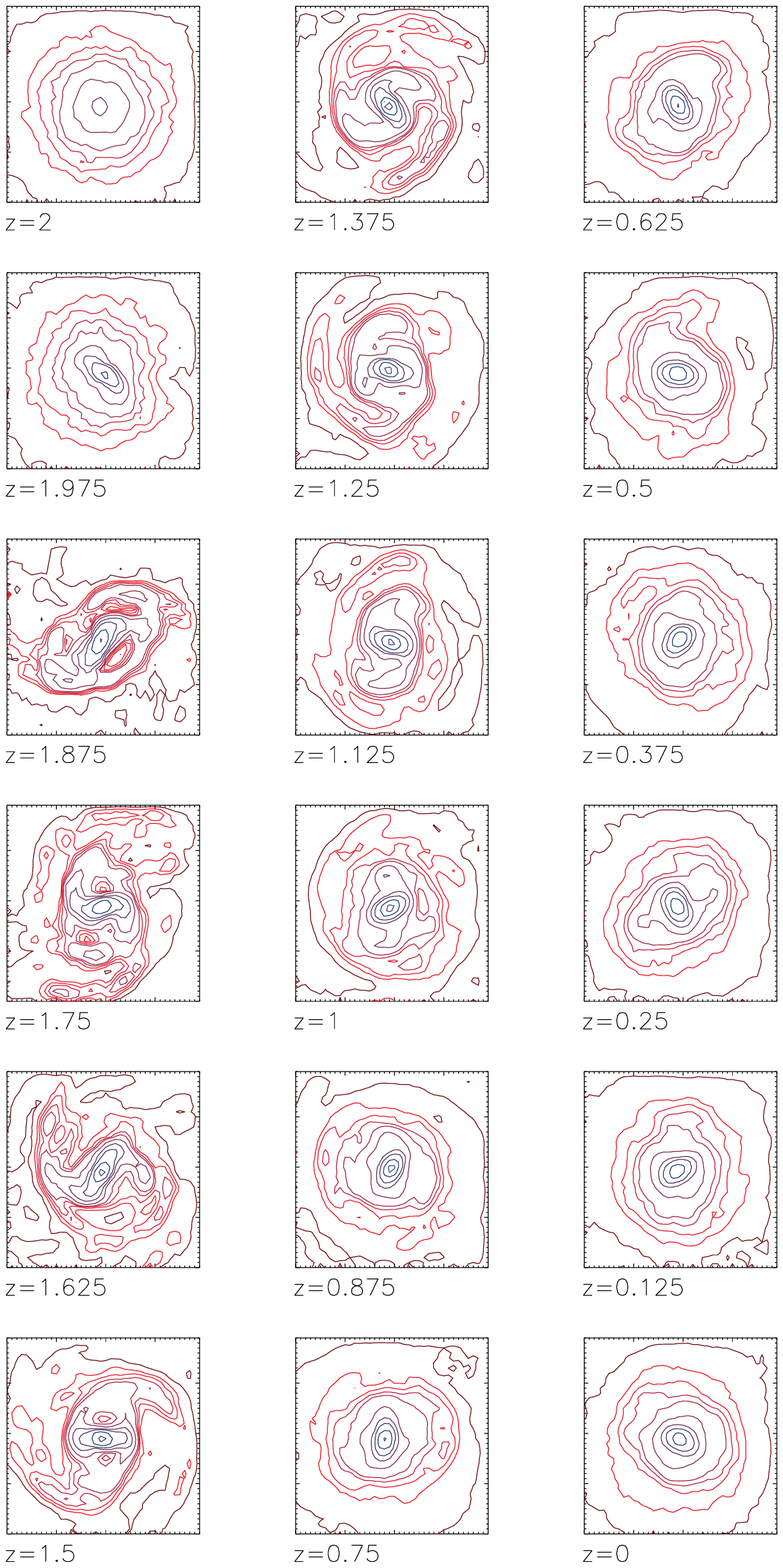}
\caption{Evolution of  isodensity contours of  simulation c3 as described in 
Fig. \ref{dens1}}
\label{dens3}
\end{figure*}

\begin{figure*}
\centering
\centering
\includegraphics[width=10cm]{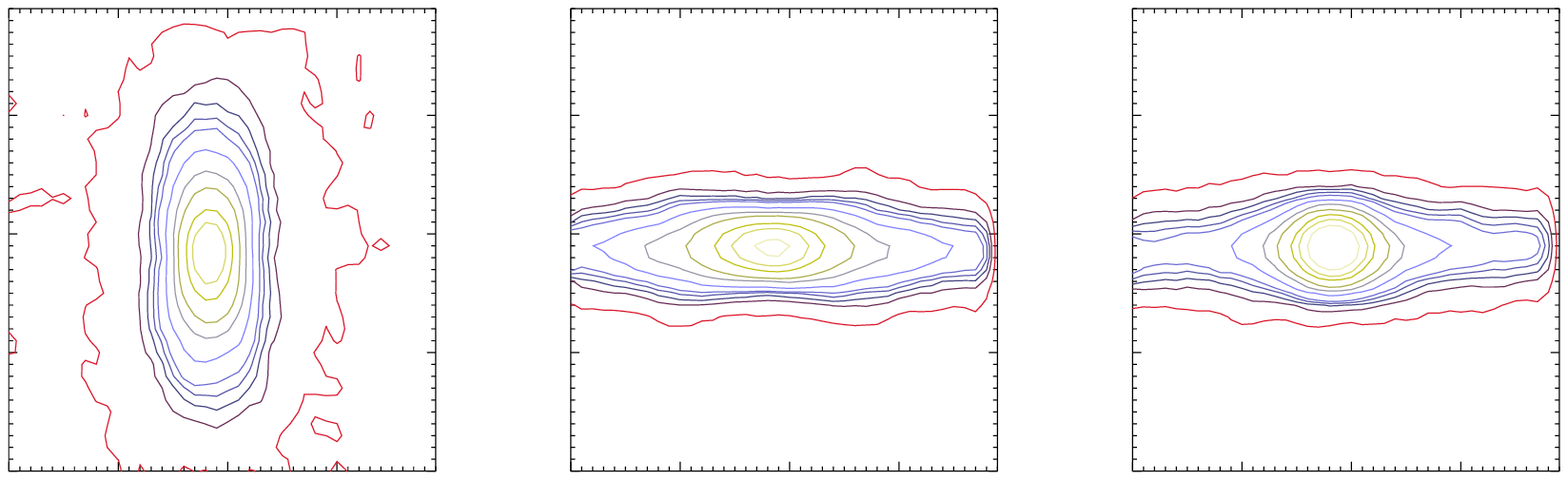}
\caption{Face-on, side-on and edge-on view of isodensity contours of simulation c1  
at $z=0$}
\label{densxz}
\end{figure*}

\begin{figure*}
\centering
\includegraphics[width=10cm]{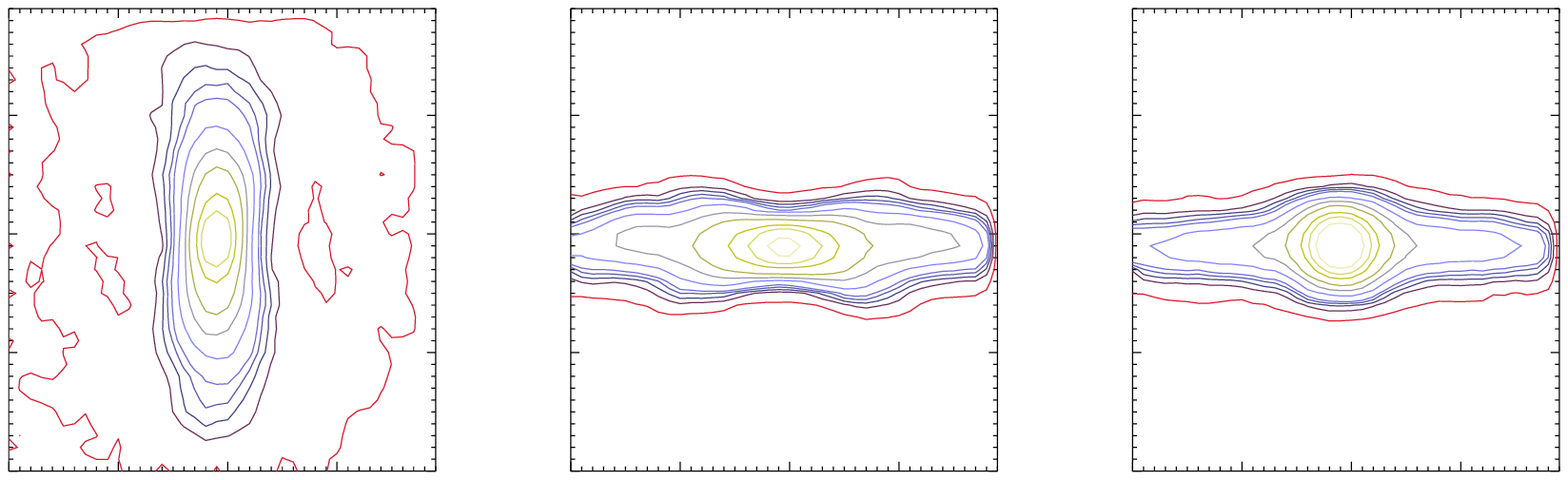}
\caption{Face-on, side-on and edge-on view of isodensity contours of simulation c2  
at $z=0$}
\label{densyz}
\end{figure*}

\begin{figure*}
\centering
\includegraphics[width=10cm]{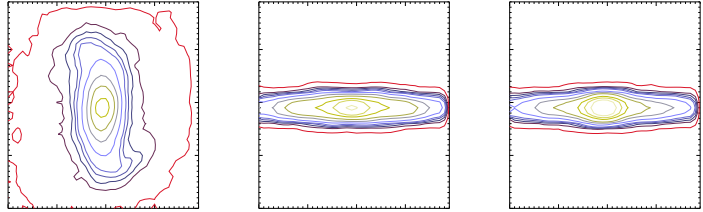}
\caption{Face-on, side-on and edge-on isodensity contours at $z=0$ of simulation
 c4 in Table \ref{cosmsimtable}.}
\label{densz0}
\end{figure*}

\begin{figure*}
\centering
\includegraphics[width=10cm]{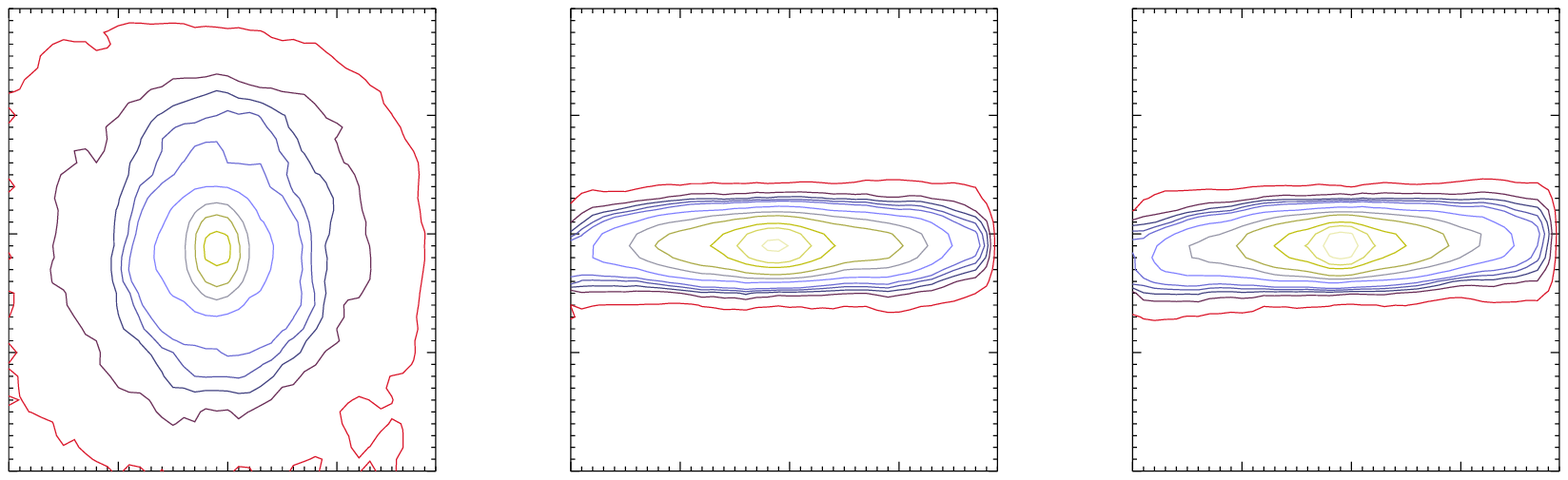}
\includegraphics[width=10cm]{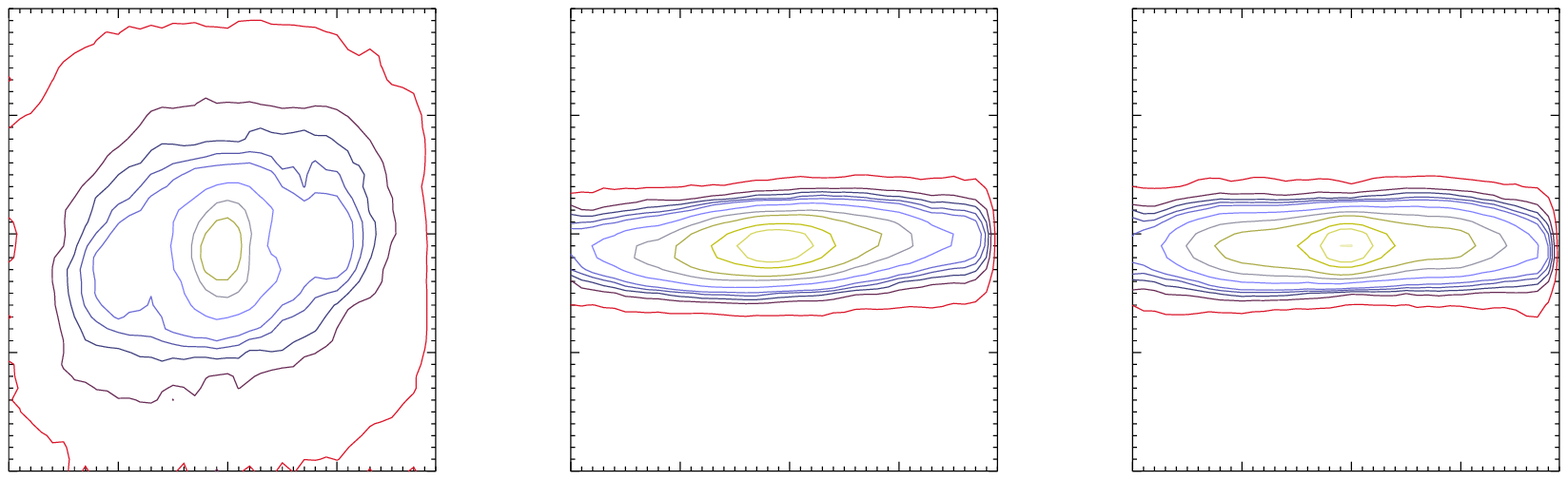}
\caption{Face-on, side-on and edge-on isodensity contours  at $z=0$ of 
simulation c3 (top panels) and  c5 (bottom panels) in Table \ref{cosmsimtable}.}
\label{densz00}
\end{figure*}

\begin{figure*}
\centering
\includegraphics[width=10cm]{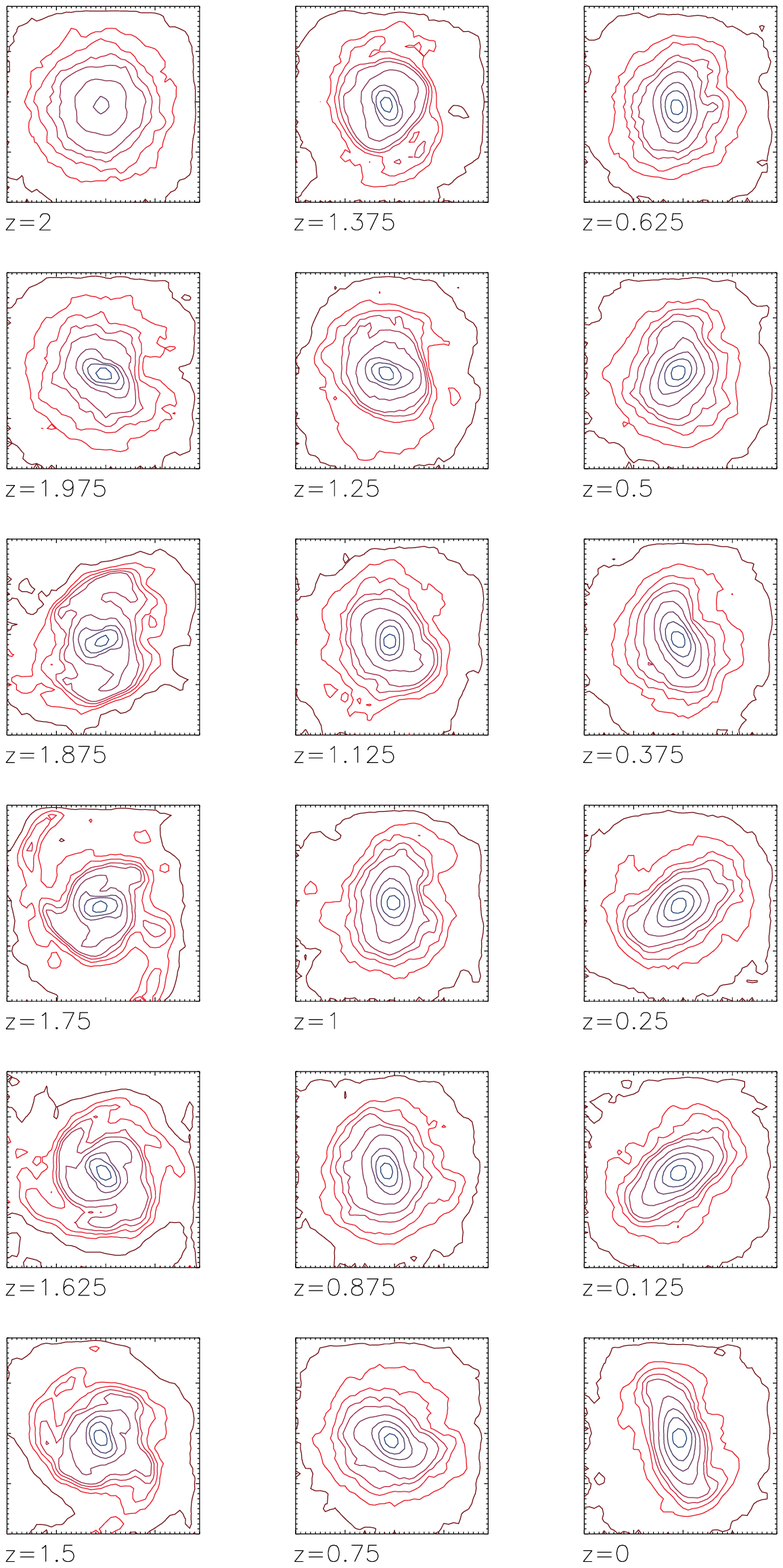}
\caption{Evolution of isodensity contours of  simulation c4 as described in
Fig. \ref{dens1}}
\label{dens4}
\end{figure*}

Fig.s \ref{dens1}, \ref{dens2} and \ref{dens3} show the evolution of 
isodensity contours of simulations c1, c2 and c3.
More massive {\it cold} disks suffer stronger lopsided instability (m=1) from  the beginning of their evolution which degenerates in the m=2
instability, i.e. the bar instability, later on.
The less massive {\it cold} disks, i. e.  DM dominated cases, show a weaker m=1
instability. So  bar instability develops before than in the
corresponding more massive cases and the disk
attempts to re-arrange before the end of the simulation. 
Fig.s \ref{densxz} and \ref{densyz}   compare  
face-on, side-on and edge-on isodensity contours  of simulations c1 and c2 at $z=0$.
We point out that
our {\it cold} intermediate  mass case  shows {\it peanut-shape} in the 
side-on view  and {\it bulge-like} contours in the edge-on view. 
Therefore, in this case,  a bulge could be mis-identified due to the bar feature. 
However in the {\it warm} analogous case (Fig.\ref{densz0}) this feature 
does not arise. On the other hand, our  more massive {\it cold} disk shows 
edge-on isodensity contours with a less defined inner bulge and quite ticker   
disk-like contours in the outer regions.
Its side-on view corresponds to a boxy image without an extreme peanut
feature.\\
Therefore the halo-to-disk ratio has a significant influence on the stellar
disk at $z=0$,\\
The {\it Q} parameter does not influence the final ($z=0$) morphologies of our
less massive disks, showing always disk-like shapes
(Fig.\ref{densz00}).
This  suggests to regard the {\it cold} intermediate mass case as a 
peculiar one as far as
 the peanut shape, is concerned. Such a feature has been
 recognized by  \citet{Comb81} as caused by vertical orbital resonances.
Fig. \ref{dens4} 
shows the isodensity contours of 
simulation c4. 
In this simulation the higher value of
the {\it Q} parameter stabilises the disk against
the local Jeans instability and the bar appears later than in
the corresponding {\it cold} case (simulation c2).\\    
Therefore {\it warmer} disks  are more stable against lopsided 
instability than  the corresponding {\it cold} cases. 
Inside {\it warmer} and  less massive disks, bars in bars, 
namely bar features at different isodensity levels, nested with twisting 
major axes, are also shown.    
\begin{figure*}
\centering
\includegraphics[width=10cm]{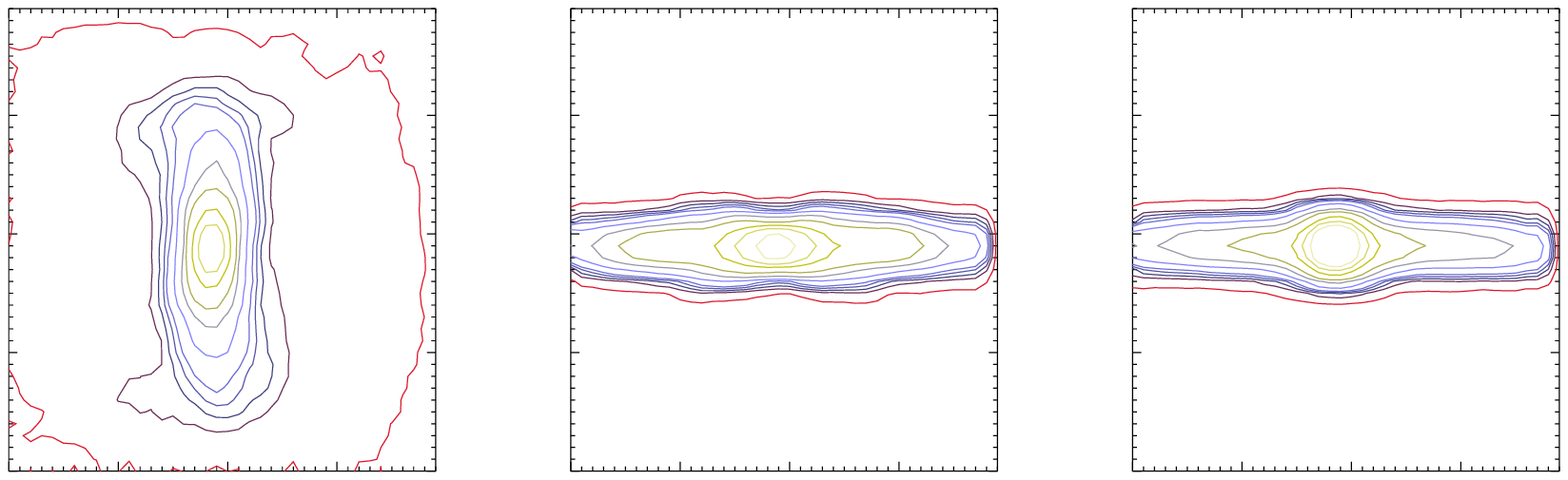}
\caption{Face on, side-on and edge-on isodensity contours of simulation c6 
at $z=0$}
\label{dens6}
\end{figure*} 
    
Isodensity contours of simulation c6 (Fig. \ref{dens6}), 
are quite similar to those  of the corresponding simulation c2 
(Fig. \ref{densyz}), 
which however starts at $z=2$. 
Morphologies of both  simulations c6 and c7 show  thinner disks than simulations c4 and c5
given  their shorter evolutionary time
($\approx 7.7 $ Gyr instead of $\approx 10.24 $ Gyr).
\begin{figure*}
\centering
\includegraphics[width=10cm]{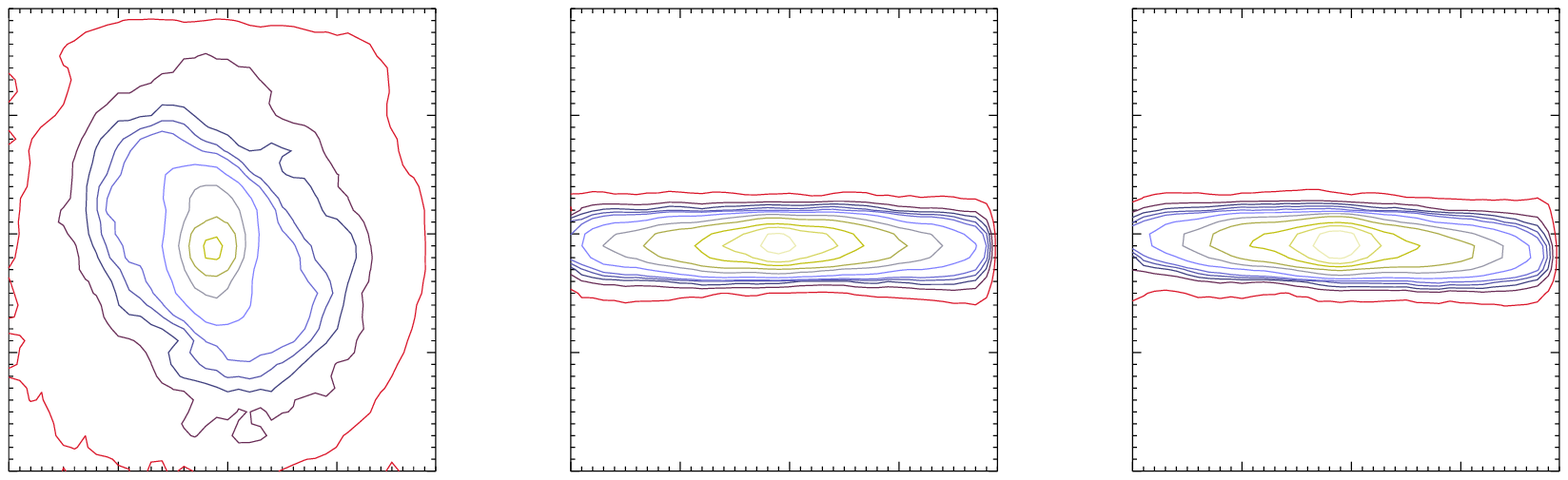}
\caption{Isodensity contours, in particular face-on, side-on and edge-on views
of the inner bar of  simulation c7 at $z=0$}
\label{dens7} 
\end{figure*}    
\subsubsection{The bar strength}
\begin{figure*}
\centering
\includegraphics[width=12cm]{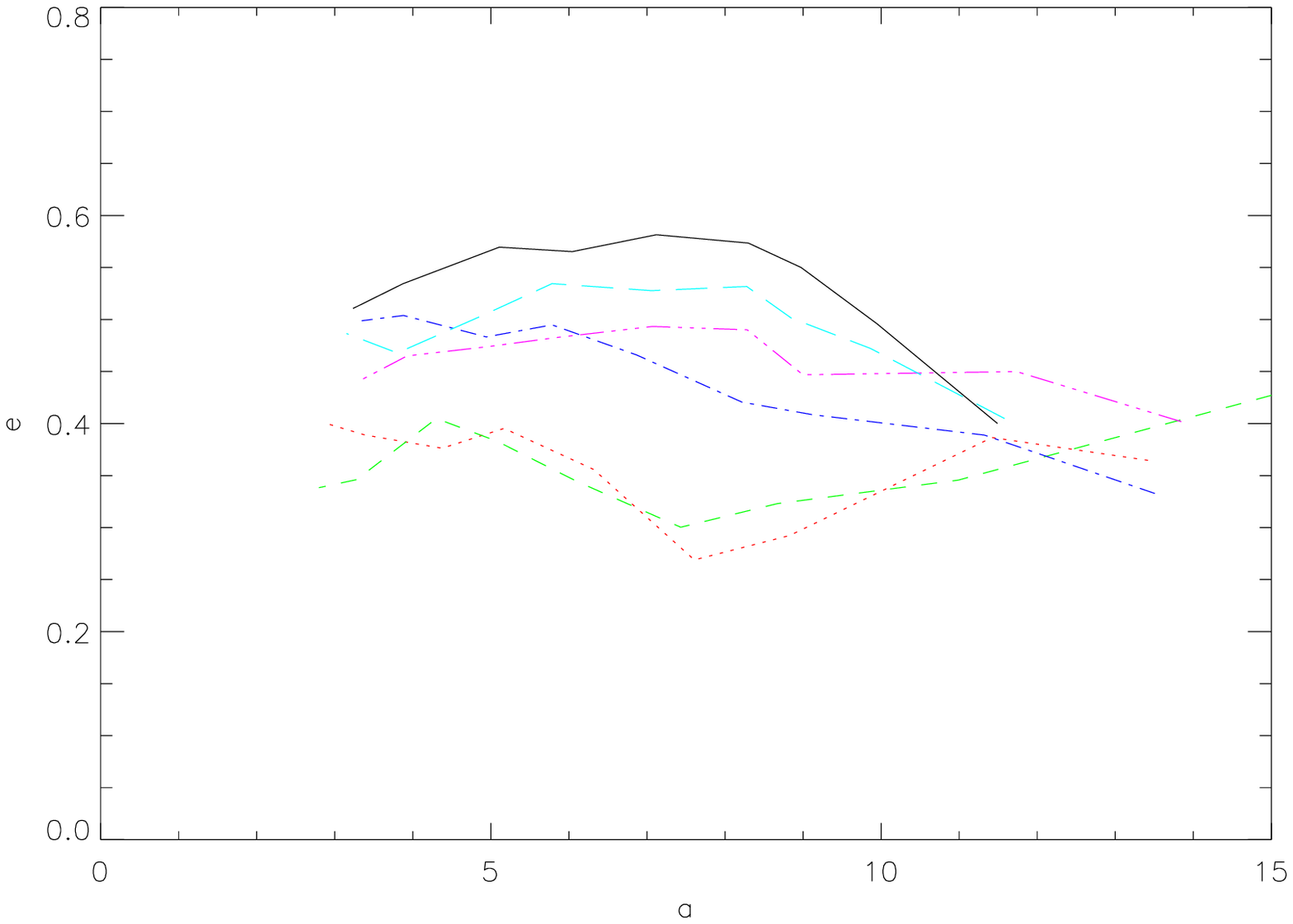}
\caption{Ellipticity as a function of the major axis, a (in physical kpc), of
  simulation c1 at different redshifts:  $z=0$ continuous line, $z=0.25$ long-dashed line, $z=0.5$ long 
and short dashed line, $z=0.75$ dot-dashed line, $z=1$ short dashed line, $z=1.25$
 dotted line. } 
\label{ell1}
\end{figure*}

\begin{figure*}
\centering
\includegraphics[width=12cm]{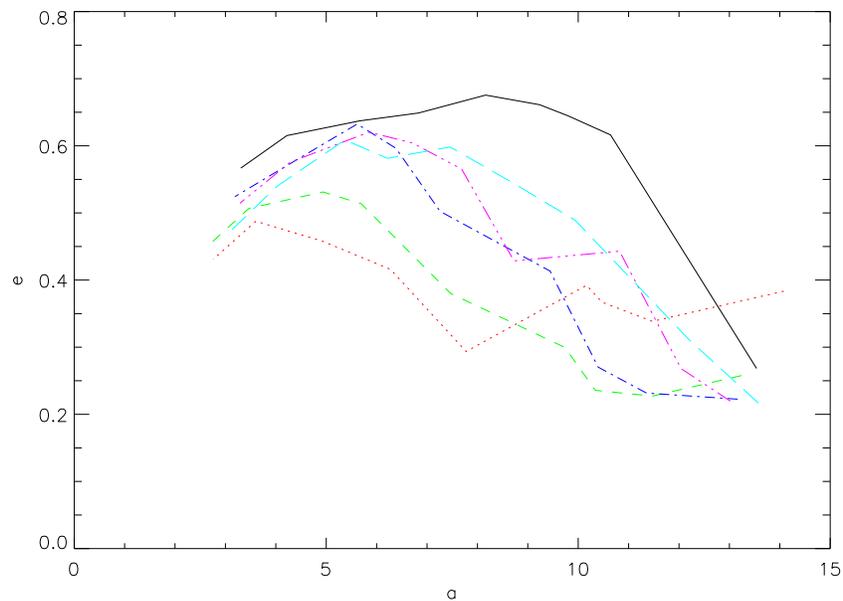}
\caption{Ellipticity as a function of major axis, a, of simulation c2
 at different redshifts; symbols are as in Fig.\ref{ell1}}
\label{ell2}
\end{figure*}

\begin{figure*}
\centering
\includegraphics[width=12cm]{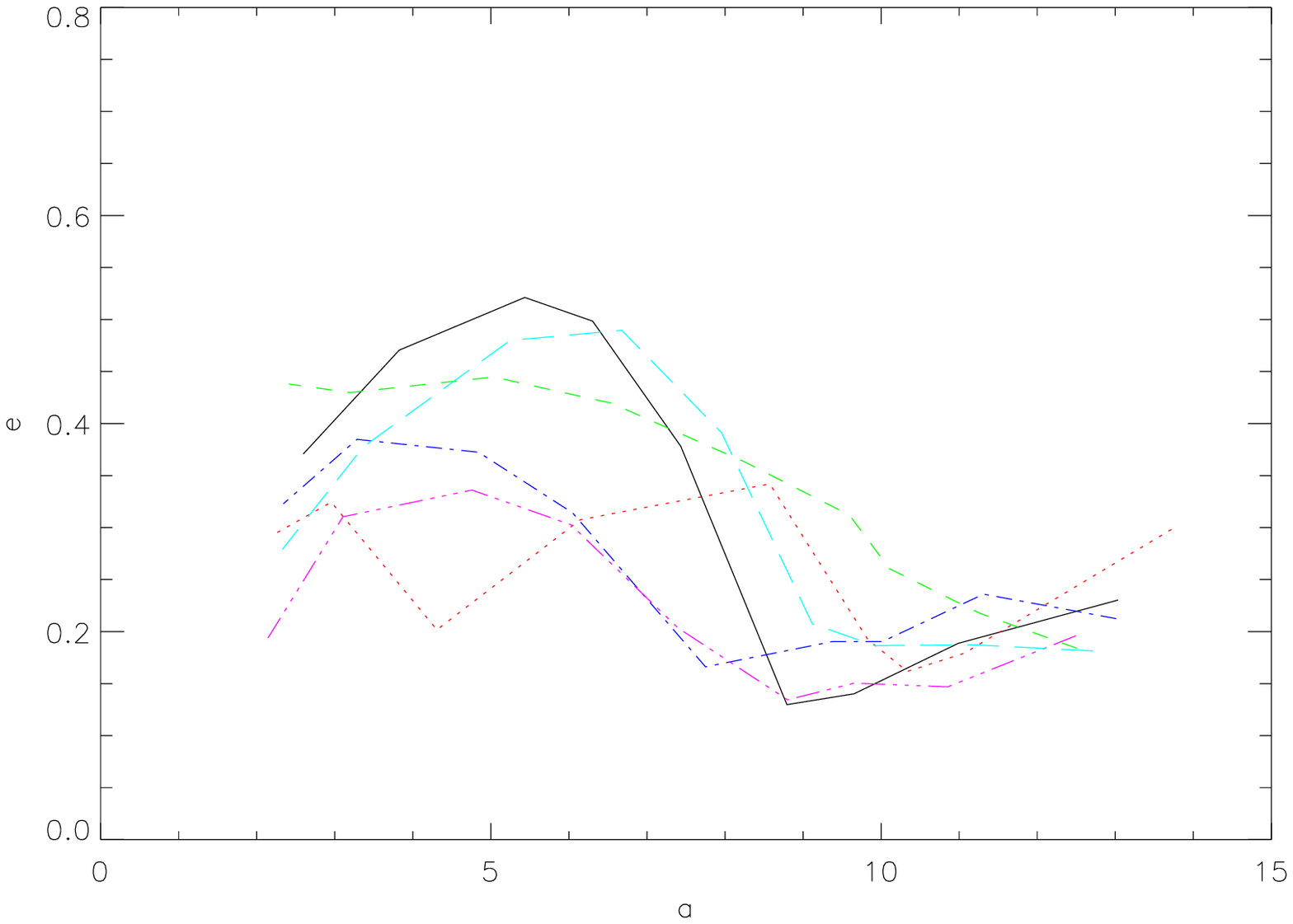}
\caption{Ellipticity as function of the major axis, a, for simulation c4 
at different redshifts; symbols are as in Fig.\ref{ell1}}
\label{ell4}
\end{figure*}

\begin{figure*}
\centering
\includegraphics[width=12cm]{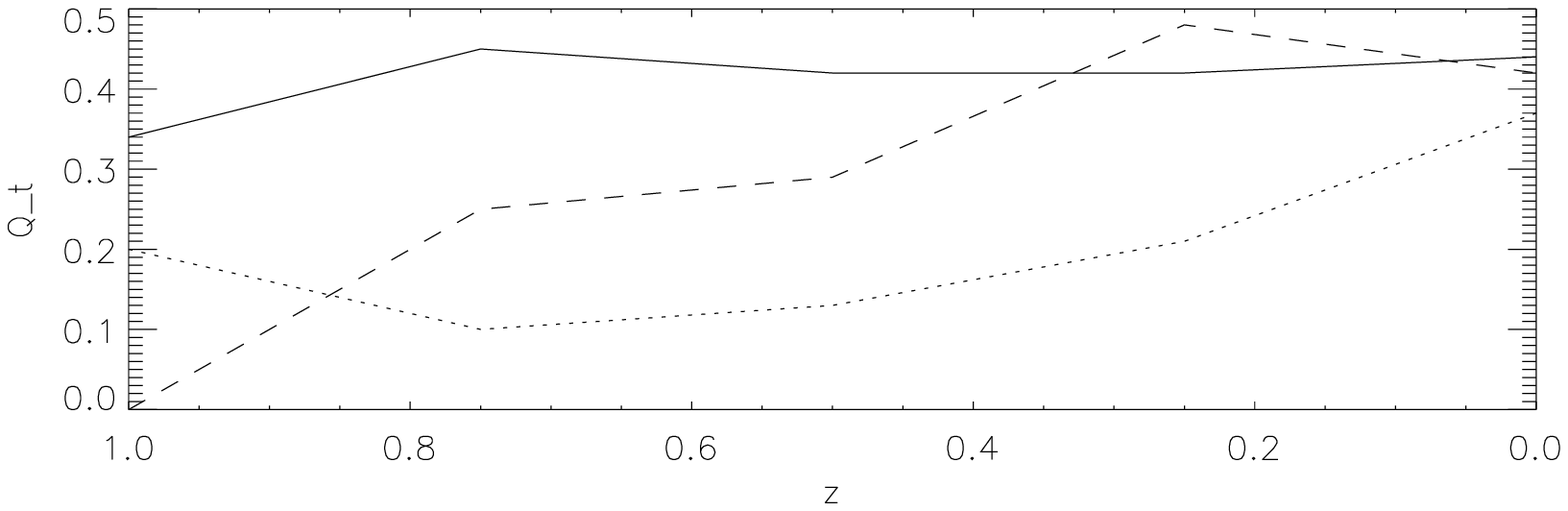}
\caption{Evolution of bar strength after $z=1$ evaluated using the
gravitational torque (see text)
 for the simulations c2 (full line), c4 (dotted line), and c6 (dashed line)  }
\label{Qg}
\end{figure*}

A variety of quantitative parameters have been suggested to evaluate  the strength of the bar (see  \citet{BuBlo01} for a review). Firstly 
we
 quantify the growth of the bar instability studying the time evolution of
the ellipticity  of our isodensity contours as a function of  their  major 
axis, $a$. The strength of the bar depends on the density contrast
accounted for, and it varies with the
distance from the centre; different choices can change its value but not the
trend outlined in Table \ref{cosmsimtable_fin}.\\
Fig. \ref{ell1} shows that in simulation c1 the strength of the bar increases 
with time.  
The length of the bar depends on the redshift too: 
it grows until $z=0.5$, and then shrinks to $z=0$.\\
By comparing Fig. \ref{ell2} and Fig. \ref{ell4}, which show the
 ellipticity profiles of 
simulations c2 and c4 respectively, 
we point out that a greater {\it Q},  in intermediate mass disks, directly 
reflects on the bar strength:
a stronger local gravitational instability, corresponding to a lower
{\it Q} value,  triggers a stronger  bar (Table \ref{cosmsimtable_fin}).
For the less massive disks, {\it Q} is poorly influent 
on the bar strength. In these cases, the local Jeans
instability has a small impact on the bar formation and evolution,
which is dominated instead by the dynamics of the DM halo.\\
The embedding redshift 
does not have a major impact on the bar strength.
Its more important effect  is the change  of the bar {\it length}  which can be
connected with the larger time span of simulation c2 (or c3) with respect to
simulation c6 (or c7).
Therefore, the halo evolution between $z=2$ and $z=1$ does not 
seriously affect the disk instability, at least for the {\it cold} disk cases.\\
\citet{Comb81} have defined the bar strength at radius R
by using  the parameter:
$Q_t= {F_T^{max}(R)\over{<F_R(R)>}}$
where $F_T^{max}=[{\partial \Phi(R,\theta)}/{\partial \theta}]_{max}$ is 
 the maximum amplitude of tangential force at radius R and 
 $<F_R(R)>=R({\partial \Phi_0}/{\partial r})$ is
the mean  axisymmetric radial force, at the same R, derived from the $m=0$
component of the gravitational potential.
We evaluated the components of the gravitational force on a
suitable two dimensional grid using the method described  by \citet{BuBlo01}.
However information provided by such approach could be affected by
spiral arm torques and by some asymmetry in the bar itself \citep{BuBlo01}. 
Nevertheless, we  succeeded in monitoring the behaviour of such a parameter 
for simulations c2, c4 and c6 (Fig. \ref{Qg}). 
The {\it{cold}} cases end up with
almost the same value of $Q_t$ even if their evolution starts from different 
redshifts. The {\it{warmer}} case, instead, maintains a smaller
value of the bar strength during all the evolution, in agreement with results 
obtained by using ellipticity parameter.
Table \ref{cosmsimtable_fin} shows that  the final values (i.e. at $z=0$) of the 
bar  strength evaluated with both these methods  are  consistent. 
According to  the  classification of
\citet{BuBlo01}, we assign class 1 to our less massive barred galaxies  
if their evolution starts from $z=2$ (i.e. simulations c3 and c5), 
class 2 if they evolve from $z=1$ (i.e. simulation c7),
and class 4  to all the other ones (i.e. simulations c1, c2, c4 and c6).
\subsection{Comparisons with  isolated cases}
\begin{figure*}
\centering
\includegraphics[width=12cm]{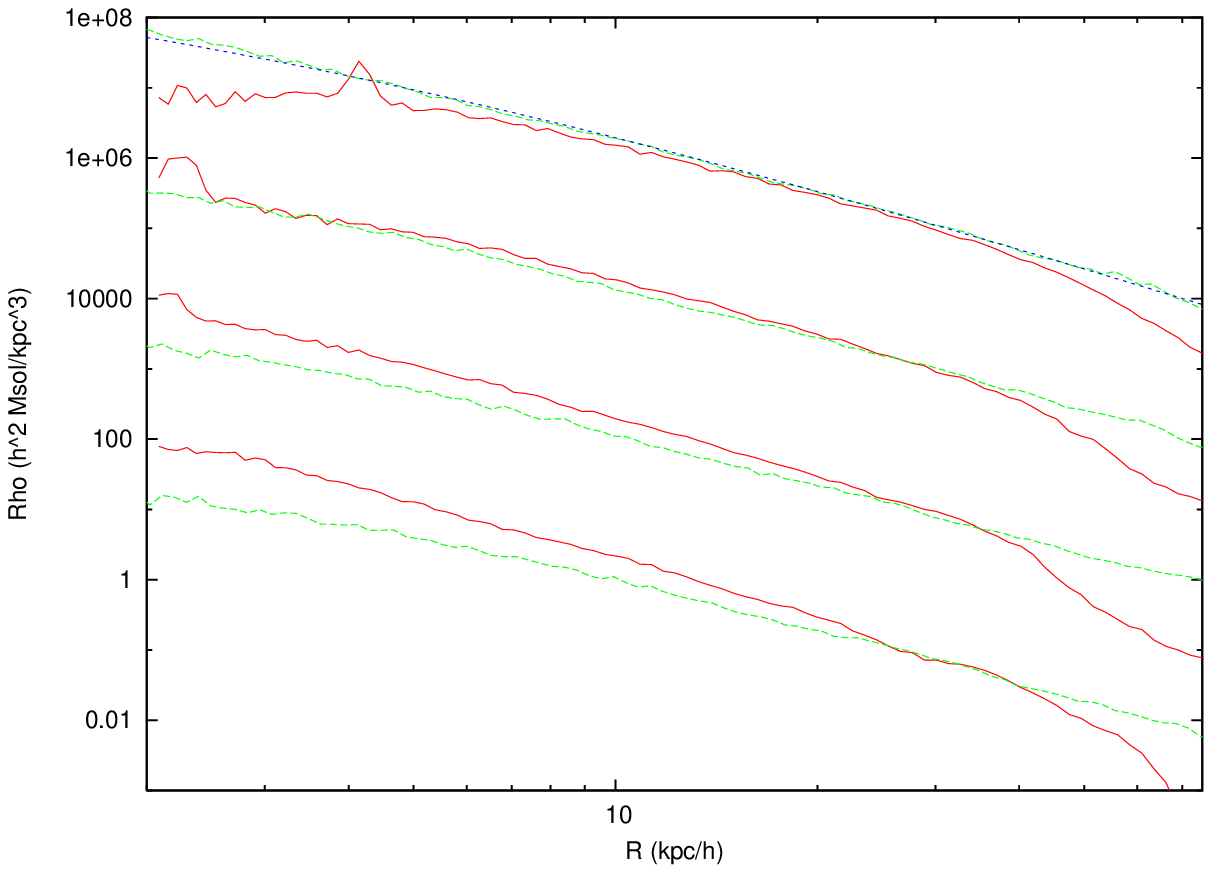}
\caption{ Radial density profiles of the DM halo in simulations c5 (dashed
lines) and i2 (solid lines) at
redshifts $z=0, 0.5, 1.0, 1.5$ from top to bottom for simulation c5 and at the
equivalent evolutionary times for simulation i2. The  three
couples of profiles below have been divided by $10^2, 10^4, 10^6$ for
clarity. We also show a NFW density profile having a concentration
parameter $c \approx 23$ (dotted line), obtained as a two parameters
best--fit of the density profile  of simulation  c5 at
$z=0$. Length units are in physical kpc.}
\label{profcomp}
\end{figure*}

In order to investigate  the  role of the   
cosmological framework  on the bar instability, we perform isolated
simulations using the same halo and the same disk-to-halo mass ratios as in our cosmological
setting (Sect. 4.2). We plug Q=1.5 as stability parameter in the 
disk.  Our results show that the less massive disks 
do not show  important differences as far as the bar 
feature is concerned:  both the bar strength is the same and  the
same bar in bar features arise.
In Fig. \ref{profcomp} we compare the halo radial density profiles of
simulations c5 and i2. The density of the halo
evolving in isolation becomes initially steeper, then it gradually flattens in
the centre. In the outer regions, where the support of the cosmological
environment is now lacking, the halo is slowly losing matter toward bigger
scales and the profile is steadily steepening.  
On the other hand, the halo evolving in the cosmological environment,
continues to accrete mass and small substructures from larger scales. Such
accretion is still significant up to redshift  $z \approx 0.5$ at least 
(Fig.  \ref{accrhist}).
Even if the dynamical evolution of the halo is different in
cosmological and isolated simulations,  the bar
in the disk does form and evolves in a similar way.
Thus we make the hypothesis that the common features of the two numerical 
experiments,
namely the dynamical evolution and the anisotropy of the mass distribution,
are the main engine for the bar instability.
The large scale cosmological environment becomes a second order effect
in the less massive disks. However the material accreting on the halo, 
which has been cut off with the 
halo segregation in a isolated system, plays a crucial role on the
degree of the disk instability if 
the disk is not  completely DM  dominated.
We conclude that the use of isolated halos in gravitational equilibrium for
the study of the bar instability can give misleading results.\\
Taking into account our previous works in such isolated non cosmological framework
(\citet{Cu99}; \citet{Ma01}), we derive that live {\it unrelaxed} halos correspond to the
most ''realistic'' approach available to simplify the picture. Even if the
caveat outlined above cannot be forgotten,
the dynamical state of the halo, as outlined in our 
works for the first time, plays a fundamental role in triggering and fuelling 
such a instability.\\

\begin{figure*}
\centering
\includegraphics[width=10cm]{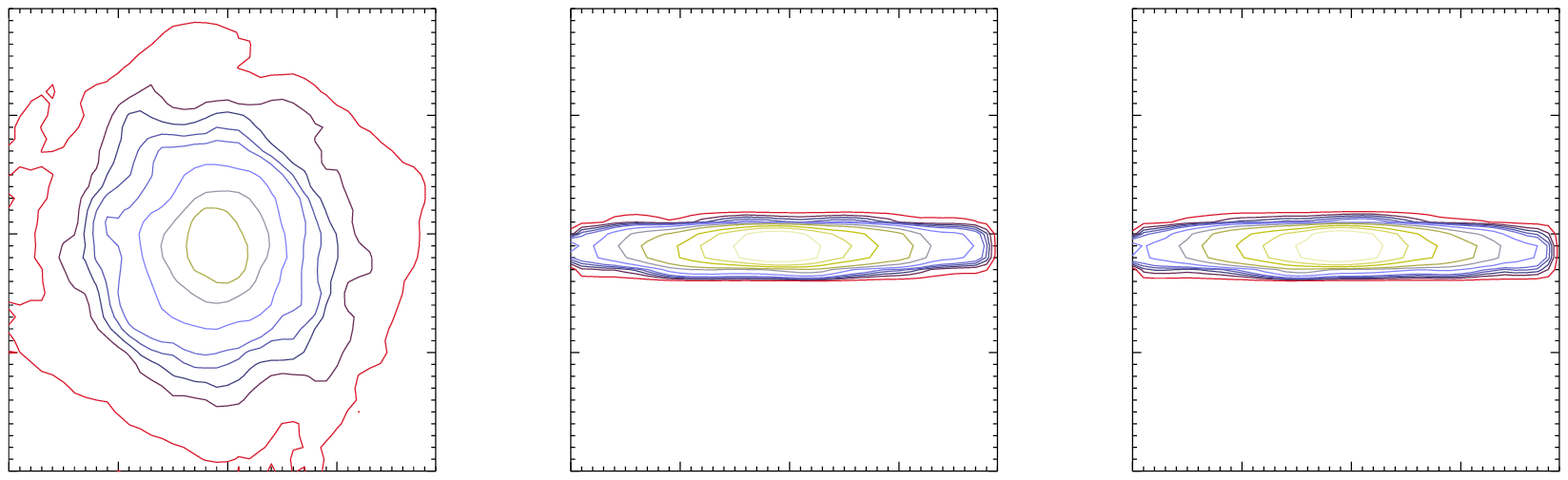}
\includegraphics[width=10cm]{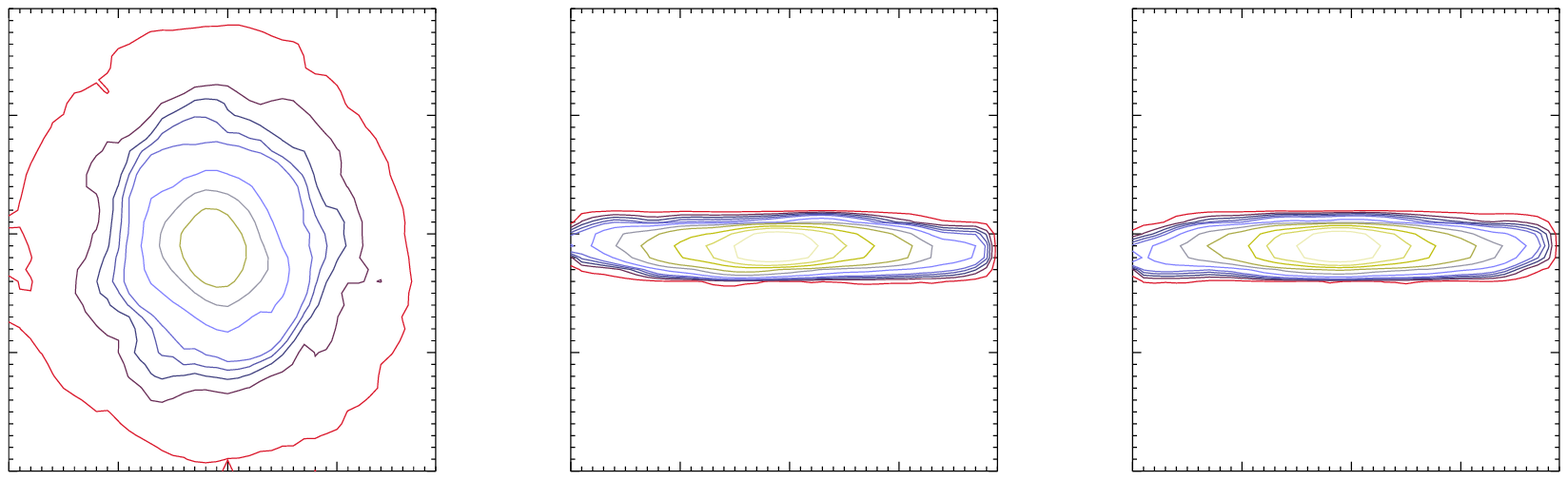}
\includegraphics[width=10cm]{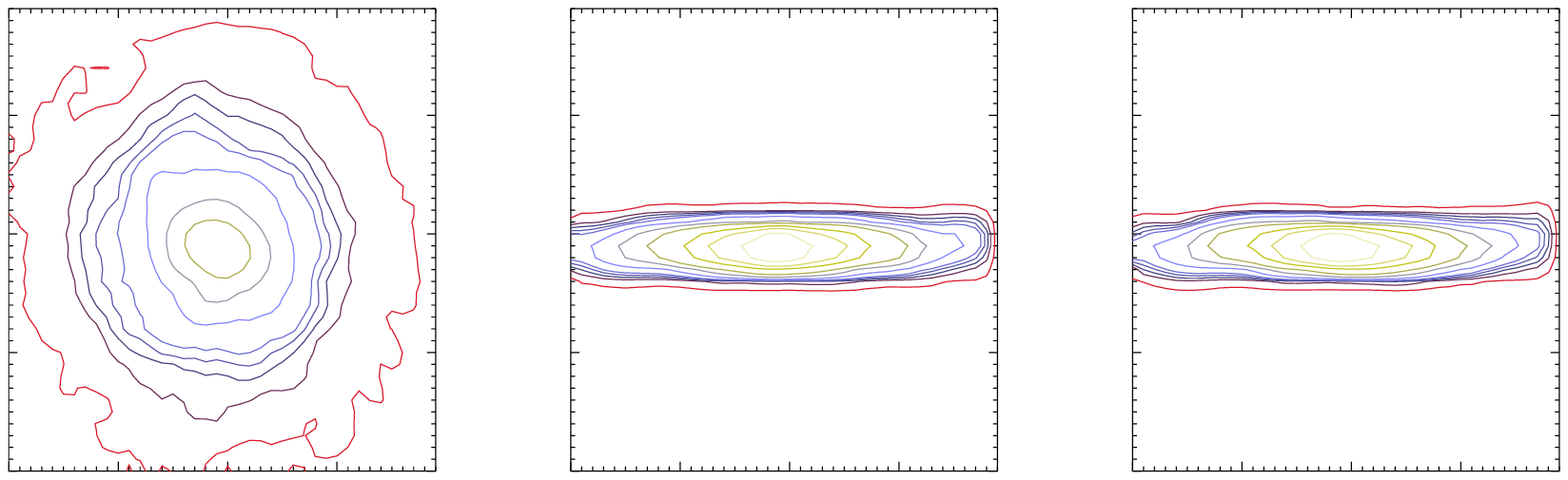}
\caption{Evolution of  isodensity contours of the simulation i5  at 
different evolutionary times: from top to bottom,    $t=5\,$Gyr, $t=7.5\,$Gyr
and $t=10\,$Gyr; xy, yz and xz projections from left to right 
(see text (Sect. 7) for more details).}
\label{NFWhalo}
\end{figure*}

In order to disentangle the role of the halo's cosmological features like the
prolate geometry and the spin
on the instability, and to test the resolution effect,
we produced an isolated halo with the same virial mass, radius and number
of particles as our cosmological halo at $z=0$, but with an isotropic NFW
radial density profile.
The procedure is described by  \citet{hern93}. 
We used a rejection technique to sample the density profile and we then assign a
velocity to each particle following a local Maxwellian velocity dispersion.
We checked that after 7 Gyr of evolution, the radial density profile of the  halo is not
changed, except for the ``evaporation'' of some particles dwelling in its outskirts.
We embedded then a disk having same mass, radius and {\it Q} as
in our simulation c5 and c3. These two simulations are labelled in Table \ref{cosmsimtable} and
Table \ref{cosmsimtable_fin} as i5 and i4.
According to the classical theory (Sect. 4.1), in simulation i5 the
bar instability would be  inhibited.  We successfully reproduced this result
with our live NFW halo (Fig. \ref{NFWhalo}). Therefore the bar instability  in
simulation c5 is a genuine effect of the cosmological evolution and there is no
evidence for a role of  numerical noise.
Moreover we note that in simulation i5  the reaction of the
DM halo to the disk immersion has {\it not} triggered a long-lived bar
instability (see Fig. \ref{NFWhalo}).
\\

\section{Discussion}
In this work  we investigate the issue of the bar 
instability in  stellar exponential disks embedded in a DM halo 
self-consistently evolving in a cosmological context. 
We aim to disentangle the effect of  few 
well-defined disk parameters on this instability. We run also isolated simulations
using the same cosmological halo to analyse the effect of the whole cosmological framework
on the results.  This paper is a re-visitation in such a cosmological scenario
of the work by \citet{Cu99}.  
To compare  our results with such paper we use the  $R_{DM}$ ratio  
in Table \ref{cosmsimtable}.
Critical threshold values against $m=1$ and $m=2$ instability for this 
ratio have been  calibrated also by \citet{atha87}. They claim for
a value around $0.81$ for the  $R_{DM}$ ratio to inhibit the lopsided instability (i.e.  $m=1$)
and around $2.2$ to suppress the $m=2$  swing amplification instability. Even if these values are derived in  a very simply framework,
i.e. a isolated spherical and analytical DM halo, they are widely used
in literature (\citet{bot03}; \citet{Elm03}),
therefore we will refer  to this parameter to analyse our initial
condition, in addition to the Efstathiou et al. parameter introduced in
Sect. 4.1. \\
Looking at Table \ref{cosmsimtable}, we point out that simulations c1, c2, c4 and 
c6, which develop strong final bars (Table \ref{cosmsimtable_fin}), are 
in the instability region for both these criteria. 
 In particular in simulation c1, which is below the
threshold of lopsided instability too \citep{atha87},  the signatures of such a 
instability are clearly shown in the
first phases of its evolution (Fig. \ref{dens1}). 
On the other hand  simulations c3, c5 and c7 are stable according to both the
criteria above.  Nevertheless a weaker bar appears and lasts until
the end of such simulations.
Therefore  the classical parameters are not  good markers 
of the onset
of the bar instability.
In particular, when the self-gravity of the disk is negligible, 
i.e. the disk is DM dominated, the halo structure generated by the cosmology
plays a crucial role in triggering such a instability. \\
Our findings  agree with results of  
\citet{May04} in the isolated framework. They found that stellar systems with disk-to-halo mass 
ratios 0.1 become bar unstable, regardless of the halo concentration
and the {\it Q} value, inside
halos built up with suitable structural parameters derived  
from  $\Lambda$-CDM cosmology,  like their circular velocity at  $R_{vir}$, $V_{vir}$, the NFW density profile and the spin parameter
(0.06 and 0.1). We point out that \citet{May04} do not take into account
cosmological evolution for their halos.
With the same disk-to-halo mass ratio \citet{atha02}  found
that such a instability is totally inhibited 
inside isotropic, non rotating halos with different density profiles
\citep[eq. 1 of][]{atha02},
in agreement   with the result of our simulation  i5 
 (Sect. 7).
This last result, together with those of simulations performed with different
number of disk particles and with different softening length (Sect. 7)
suggests that the development of long-living bars seen in our simulations is a
genuine physical effect and not a numerical artifact.\\
Bar instability in the DM dominated cases is strongly affected by the 
halo models. Moreover structural details of the halo, related to the
cosmological framework, drive  morphological features
of the stellar disk. {\citet{May04} found indeed a central bulge
after 7 Gyr which does not appear in our corresponding case. However such a bulge
shows up  in our intermediate self-gravitating case (simulation i1 in Table \ref{cosmsimtable}). 
This feature is also emphasised
in the work by \citet{athami02}) and \citet{atha03} 
for  a disk-to-halo mass ratio 0.2, using the same halo presented  in \citet{atha02} with
the higher halo concentration.\\    
Our results here are in good qualitative agreement 
with those by \citet{Cu99} concerning their simulations 3, 4, 7 and 8, with the same 
disk-to-halo mass ratio as in simulation c1 of Table \ref{cosmsimtable}, and also 
with their simulations 5 and 6
which correspond to a disk-to-halo mass ratio 0.2. 
However, we remark  that  their simulations  3 and 4 
correspond to a {\it relaxed} halo,  whereas 
5, 6, 7 and 8
to a {\it unrelaxed} halo. In particular the $R_{DM}$ initial values of their simulations
5 and 6 are respectively above and  very near  the 2.2 threshold value of bar instability, nevertheless the bar
lasts until the end of their simulations ($\simeq\,1.5$\,Gyr).
Their simulations 1 and 2, which correspond to a
{\it relaxed} dynamical state of a  halo with disk-to-halo mass ratio 0.2, emphasise however a very different behaviour as far as the 
bar instability is concerned:  the bar forms initially but  degenerates then in a 
dense nucleus. 
Thus we argue that  an {\it unrelaxed} dynamical state for 
isolated halo systems is more suitable to mimic a realistic 
{}``cosmological{}'' halo, characterised by evolution, substructure and in-fall. 
This finding is important, since a vast 
majority of the work on the bar instability  assumes a 
\emph{gravitationally stable} halo.
\section {Conclusions}
In this work  we present the first attempt to analyse the growth of
bar instability in a fully consistent cosmological framework. 
We investigate such a  issue in  stellar  disks embedded in a DM halo
self-consistently evolving in a cosmological context.
We aim to disentangle the effect of  few
well-defined disk parameters on this instability. We run also isolated simulations
using the same cosmological halo to analyse the effect of the  cosmological framework.
Our results show that:\begin{itemize} \item{
 stellar disks of different properties, i.e. mass and {\it Q} parameter, embedded in
the same halo and evolving  in a fully consistent
cosmological scenario,  develop long living bars lasting
up to redshift 0.}
\item{
The  classical criteria to account for bar instability
cannot be validated in a cosmological framework where
 a bar always develops, due the halo
evolution.}
\item{
The strength of the bar at $z=0$ is weakly depending on the {\it Q} parameter, for a
given disk mass. However for the same disk-to-halo mass ratio, colder disks
show stronger and longer bars. Thus the less massive {\it warm} disks entail
the
weakest bars, moreover  bar in bar  is a common feature in their face-on
morphology.}
\item{
Simulations performed embedding different disks in the same halo, extracted
at $z=2$ from the cosmological framework, show that the effects of the large
scale structures are negligible in the less massive,  DM dominated disks.}
\end{itemize} 
Moreover by comparing results in this work with our previous paper \citep{Cu99},
we point out that live {\it unrelaxed} halos are the more suitable approach to mimic
cosmological halos and to analyse bar instability in the less massive disks.\\
The mass anisotropy and the dynamical evolution of the DM halo have a crucial
effect in enhancing and fuelling the bar instability, also in cases where {\it
  ad hoc} halo models provided stability predictions,
\citep[e.g.][]{atha03}.
The large--scale effects, such as the continuous matter infall on the halo and
the  infall of substructures during the whole time--span of the
simulation, influence the bar strength and the details of its structure.\\ 
{\bf Acknowledgements}  
Simulations have been performed on the CINECA IBM SP4 computer (Bo, Italy), thanks
to the INAF-CINECA grants cnato43a/inato003 ``Evolution of disk
galaxies in cosmological contexts'', and on the Linux PC Cluster of
the Osservatorio Astronomico di Torino. We wish to thank for useful
discussions: T. Abel, S. Bonometto, A. Burkert, E. D'Onghia,
F. Governato, A. Klypin \& V. Springel. 

\appendix

\section{Numerical tests}

\subsection{Initialisation procedure}
While the
disk is built to be in equilibrium with the gravitational potential
of the halo (Sect. 2), the opposite is not true.
This means that the halo does contract when we embed the disk,
the amount of such a contraction depending on the mass of the disk.
This effect roughly mimics the so--called ``adiabatic contraction''
\citep[e.g][]{Jes02}  of a DM halo in a self--consistent
galaxy formation scenario, when the gas firstly collapses to
form a disk. Nevertheless, in our approach the disk appears suddenly and also
contracts in
reaction to the augmented DM concentration. This double
effect could trigger a bar instability.
Therefore we modified the code to allow for
the presence of a {\it frozen} stellar component. 
Its particles do produce a gravitational force on the DM ones, but
they are not subjected to any force.
We built up a isolated frozen disk+halo system as described above (Sect. 4.2).
Each particle of the disk is initially massless; such a 
mass increases linearly with the time until a final disk-to-halo mass
ratio  $0.3$. The system has been integrated in physical coordinates until
$T=0.51$ Gyr, which corresponds to $\approx 1.5\,t_{dyn}$ (Sect. 4.2). 
After this period, {\it a new disk having
the same mass ratio  is embedded, in equilibrium with the gravitational
potential of the halo}. We evolved this
system up to the final time, $T=10.24$ Gyr. This simulation is presented in
Table \ref{cosmsimtable} and Table \ref{cosmsimtable_fin} as simulation i3.
The presence of a very
strong bar is  seen during all the evolution. 
In Fig.  \ref{isolfroz} we compare  the evolution of the
maximum bar ellipticity and of the corresponding semi-major axis length for  
simulation i3 and i1, in which the disk was added impulsively. 
\begin{figure*}
\centering
\includegraphics[width=12cm]{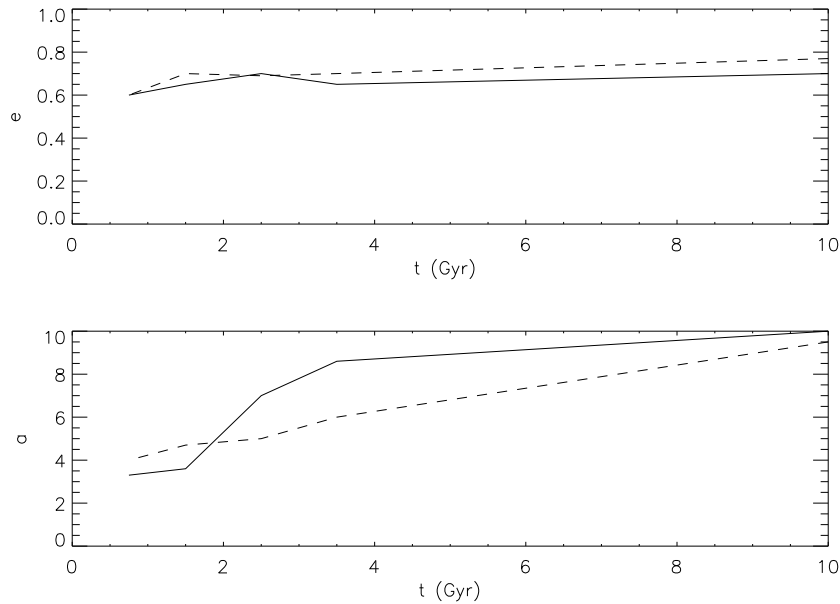}
\caption{Evolution of the ellipticity (top panel) and of the bar length (bottom panel) 
of simulations  i1 (dashed line) and i3 (full line) }
\label{isolfroz}
\end{figure*} 
We conclude that the 
reaction of the DM halo to the disk immersion, and the following 
counter-reaction 
of the disk to the increased halo concentration, are not the most important
ingredient in producing the   long--lasting bar we observe.\\
\subsection{The mass resolution}

Using a fixed number of star particles,
their mass varies by  a factor of 10 in our simulations (Table
\ref{cosmsimtable}). This can bring to a
different numerical scatter between star and DM particles. 
Problems connected with the resolution could be the
occurrence of a bar instability triggered by numerical noise, which causes a poor 
sampling of the density field in the central part of the halo, and
the numerical heating, which also depends on the number of particles involved
\citep{Lacey85}, could artificially dump the bar instability. 
Therefore we run isolated tests with different  number of star
particles ($N$= 9400,18800, 56000, 560000). We outline that results are  slightly
depending on the mass resolution of disk particles (Fig. \ref{numpart}).  
In particular also the disk with 10 times more particles than our fiducial
case keeps the bar feature, showing  a more defined bar structure due to
the higher resolution.  
We conclude that,
even if there is  a slight dependence of the details of the bar structure on
the mass resolution of star particles, our main result, i.e. the
occurrence of a long lasting bar instability even with this disk-to-halo mass
ratio which would be classically stable, it is not caused by a 
resolution effect.\\  

We are currently performing new cosmological simulations (as in Sect. 4.1) 
with higher mass resolution: 
a new halo consisting  of 282134 particles at $z=2$ (compare with Table \ref{halotable}) 
and  a stellar disk endowed with 280000 particles. The disk-to-halo mass
ratio, defined as in Sect. 4, is $1/10$. The halo has a spin
$\lambda$=0.026, a  viral mass of \(7.33\cdot 10^{11}h^{-1} \) M\(
_{\odot } \) at $z=0$  and a ``quiet'' mass accretion history, similar to
that of the DM halo used in this work.
Using such halo, evolving in a different cosmological environment, 
we  aim not only to improve the mass resolution of  simulations in the  
cosmological framework but also 
to check how our results are depending on the properties of the  halo.
Fig. \ref{hires}  shows that a bar is present  
for such a very light disk  also  with the higher resolution.\\
\begin{figure*}
\centering
\includegraphics[width=10cm]{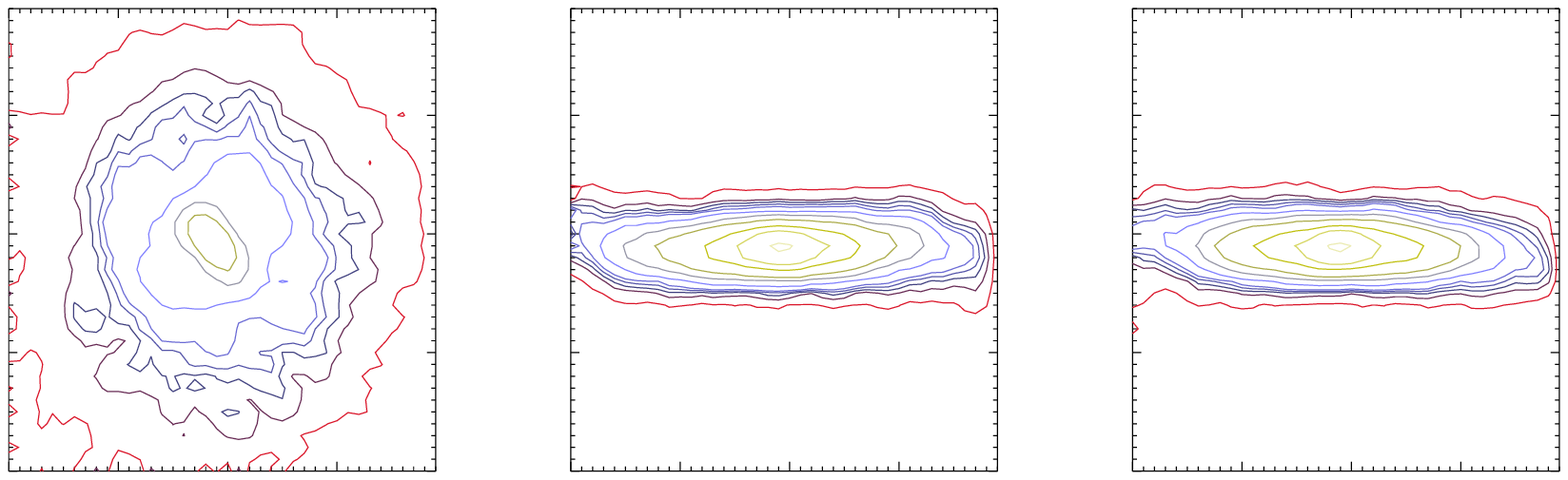}
\includegraphics[width=10cm]{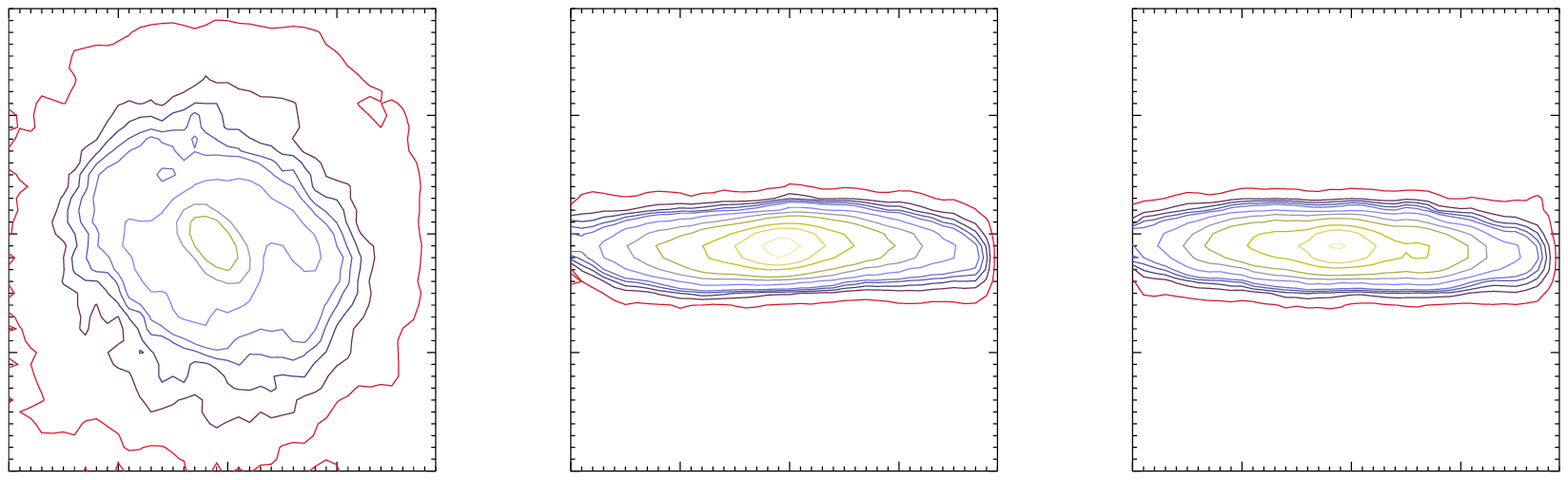}
\includegraphics[width=10cm]{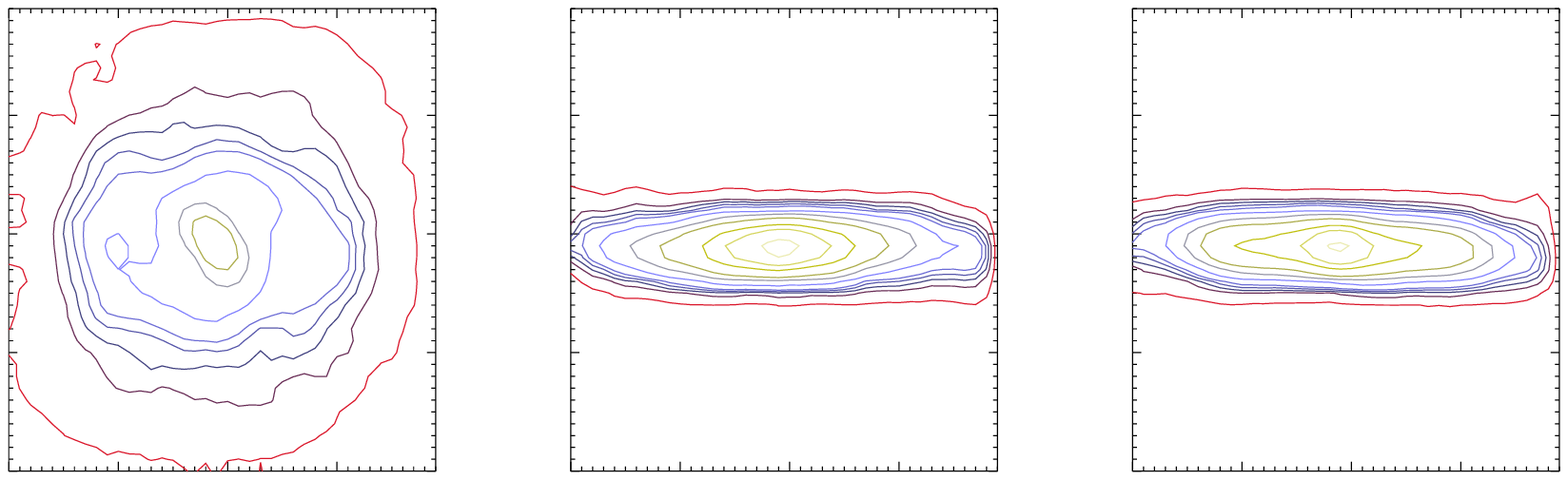}
\includegraphics[width=10cm]{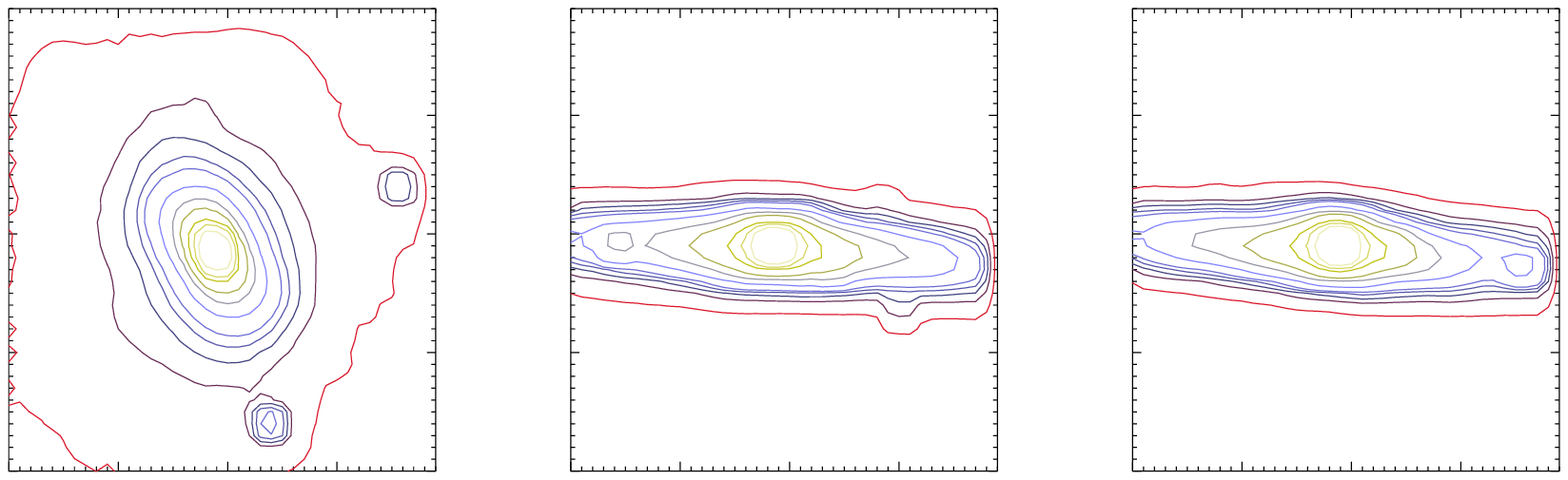}
\caption{Isodensity contours at $T=7.0\,Gyr$ of isolated simulations i2  with
different number of star particles, 9400, 18800, 56000 and 560000 from top to
bottom; xy, yz and xz projections are from left to right.}
\label{numpart}
\end{figure*}

\begin{figure*}
\centering
\includegraphics[width=10cm]{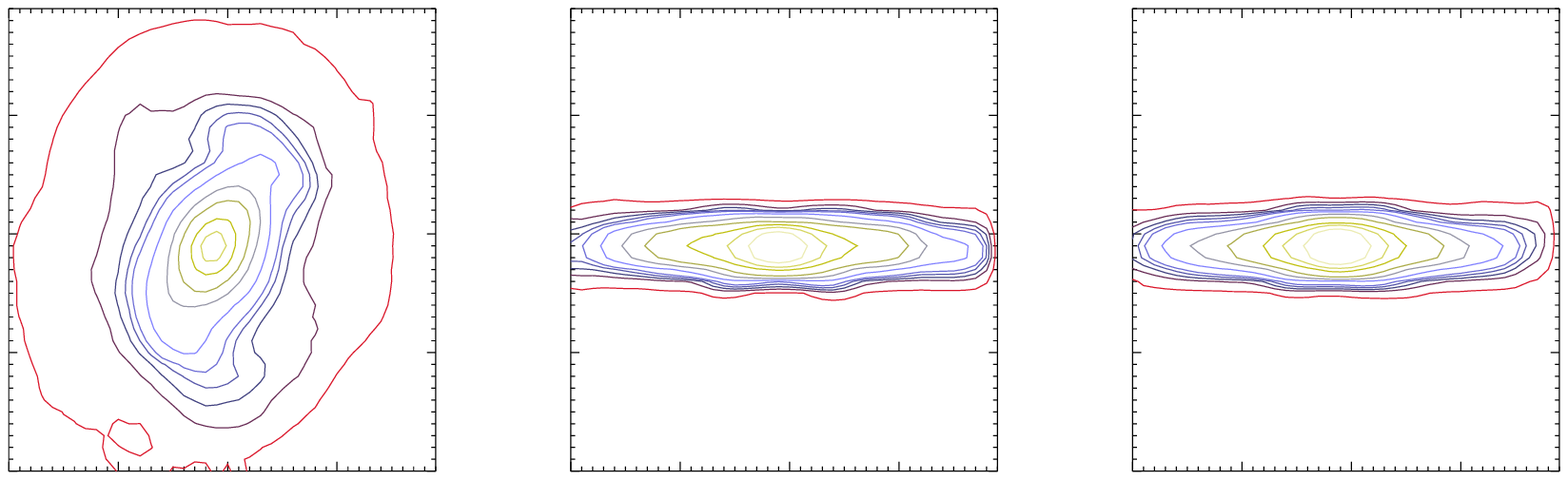}
\caption{ xy, yz and xz projections from left to right after 7 Gyr for the
high resolution cosmological run with disk-to-halo mass ratio 0.1. 
See text (Sect. 7) for more details.}
\label{hires}
\end{figure*}

\subsection{The softening length}
Our fiducial softening is similar to the one suggested by \citet{Pow03}  in their numerical convergence analysis of
the radial density profiles of DM halos. Accounting for our number of
particles, they suggest a minimum softening length:
$\epsilon_{min} \approx R_{200}/\sqrt N_{200}$, to prevent strong
discreteness effects. From Table \ref{halotable}, this value corresponds
to $\epsilon_{min}=0.33 h^{-1}$ kpc at $z=0$ and
$\epsilon_{min}=0.15 h^{-1}$ kpc at $z=2$. 
Thus we run three simulations of the
same isolated disk+halo system as in simulation i2 using three different 
softening lengths: 0.36$h^{-1}$, 0.5$h^{-1}$ and 
0.65$h^{-1}$\,kpc.  In Fig.\ref{soft} we show the isodensity contours  
at times corresponding to  redshifts $z=1$ and $z=0$ in the cosmological
framework. The bar shapes and
strengths are  similar at $z=1$, where comparable inner bars (as given by two inner
levels) are shown. We note the same similarities at $z=0$ with the exception of
simulation with lower softening length, which shows more round inner contours.
We conclude
that, as far as the bar instability is concerned, varying the stellar
softening within such range has {\it not} a significant effect on the
results.\\
\begin{figure*}
\centering
\includegraphics[width=10cm]{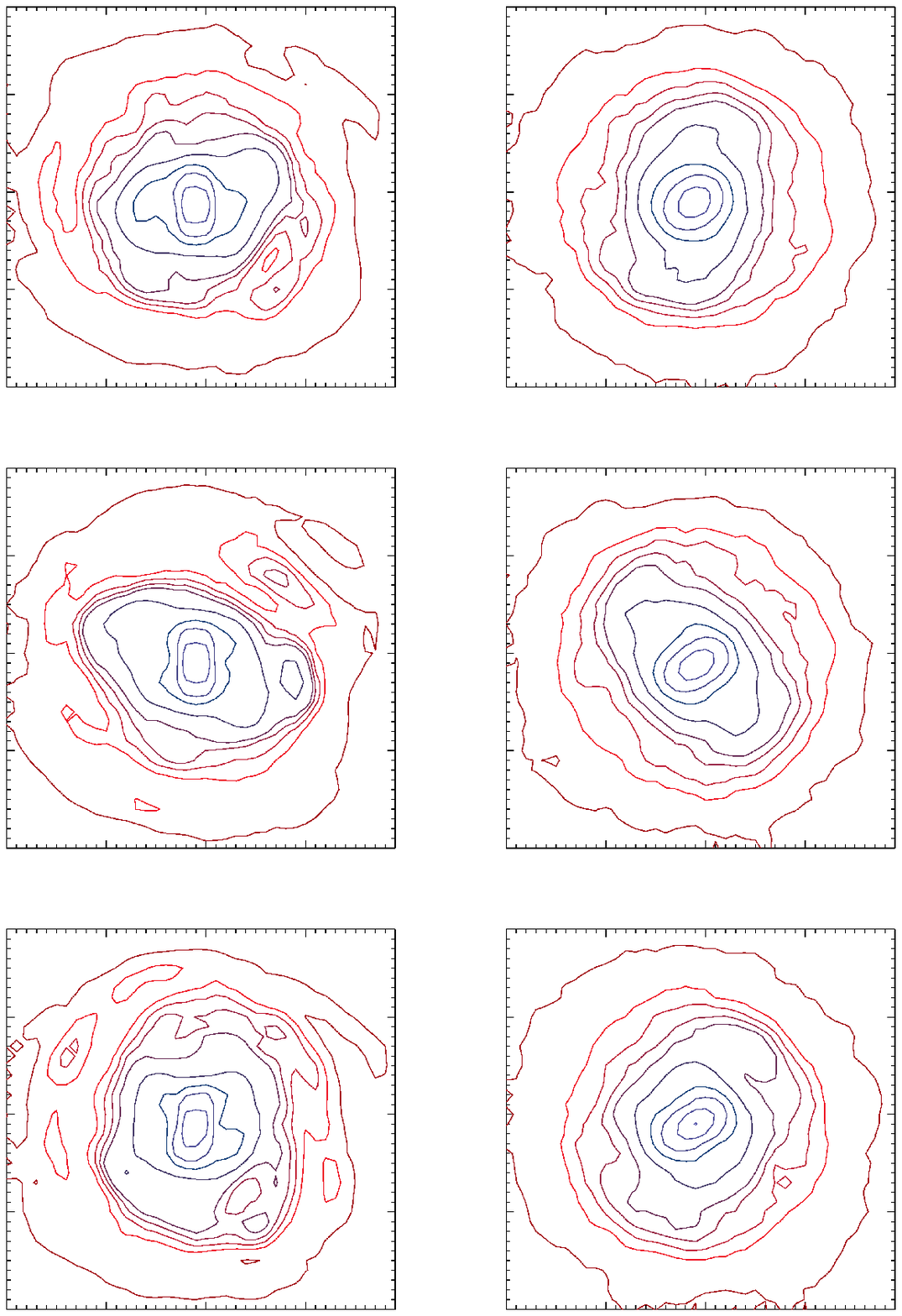}
\caption{Face-on views at T=7.71\,Gyr and 10.24\,Gyr, corresponding to $z=1$ 
(left) and $z=0$ (right) in the cosmological framework, 
of isolated simulations i2 with stellar softening length
0.36 $h^{-1}$ , 0.5  $h^{-1}$ and 0.65  $h^{-1}$ \, kpc respectively from top to bottom (see text 
for more details)}
\label{soft}
\end{figure*}
\subsection{The disk radius}
The last simulation we performed was devoted to explore if the radius of the
stellar disk can affect the triggering  of  bar instability. Therefore we
carried out a simulation with disk parameters as in simulation i1,   but with
both disk radius and scale length smaller by  a factor of two. 
Results do not differ from the comparison case: the system  shows a definite
bar instability developing after 0.25 Gyr and lasting until the end of the
simulation.

\bibliographystyle{aa}
\end{document}